\documentclass[11pt,a4paper,dvips]{article}
\usepackage{a4p}
\usepackage{rotating}
\usepackage{graphicx}
\usepackage{subfigure}
\usepackage{longtable}
\usepackage{multirow}
\usepackage{textcomp}
\usepackage{authblk,colortbl}
\usepackage{units}
\usepackage{amsmath,amssymb,amsfonts} 
\usepackage{mathrsfs}
\usepackage{hyperref}

\def\beq{\begin{equation}} 
\def\eeq{\end{equation}}
\def\bea{\begin{eqnarray}}
\def\eea{\end{eqnarray}}
\def\eqref#1{eq.~(\ref{eq:#1})}

\def\nn{\nonumber}   

\begin{document}

\begin{titlepage}

\begin{flushright}
{ \bf IFJPAN-VII-2025-26  \\
}
\end{flushright}

\vspace{1.0cm}
\begin{center}
  {\Large \bf TauSpinner algorithms for including spin}\\
  {\Large \bf and New Physics effects in $\bar q q \rightarrow Z/\gamma^* \to \tau \tau$ process}
\end{center}

\vspace{0.15cm} 
\begin{center}
  {\bf  A.Yu.~Korchin$^{a,b,c}$, E.~Richter-Was$^c$ and Z.~Was$^{d}$} 
	\vspace{0.35cm}
	
{\em  $^a$NSC Kharkiv Institute of Physics and Technology, 61108 Kharkiv, Ukraine} \\
{\em  $^b$V.N.~Karazin Kharkiv National University, 61022 Kharkiv, Ukraine} \\
{\em  $^c$Institute of Physics, Jagiellonian University,  ul.~Lojasiewicza 11, 30-348 Krakow, Poland}\\ 
{\em $^d$Institute of Nuclear Physics Polish Academy of Sciences, 31-342 Krakow, Poland}\\

\end{center}
\vspace{1.0 cm}
\begin{center}
{\bf   ABSTRACT  }   
\end{center}

The possible anomalous New Physics contributions to dipole and weak dipole moments
of the $\tau$ lepton bring renewed interest in development and revisiting charge-parity
violating signatures in $\tau$-pair production in $Z$-boson decay at energies of the LHC.
In this paper, we discuss effects of anomalous contributions to polarisation and spin correlations
in the $\bar q  q \to \tau^+ \tau^-$  production processes, with $\tau$ decays included.
Because of the complex nature of the resulting distributions, Monte Carlo techniques are useful,
in particular of event reweighing with studied New Physics phenomena. Extensions of the
Standard Model spin amplitudes, within Improved Born Approximation used for matrix element,
are implemented in the {\tt TauSpinner} program.
This is mainly with $\tau$ dipole and weak dipole moments in mind, but is applicable
to arbitrary New Physics interactions, provided they can be encapsulated into the
Standard Model $2 \to 2$ structure of matrix element extensions. Implementation allows one also to
introduce arbitrary phase-shift between vector and axial-vector couplings of $Z$ boson to $\tau$ leptons,
which would have impact on observed transverse spin correlations.
Basic formulas and algorithm principles are presented, together with distributions for spin
correlation matrix. Numerical examples of impact on  experimental signatures are shown
in case of $\tau^\pm \to \rho^\pm \nu_\tau \to \pi^\pm \pi^0 \nu_\tau$ decays. Information on 
how to use and configure the {\tt TauSpinner} program is given in Appendix.

\vfill %
\vspace{0.1 cm}

\vspace*{1mm}
\bigskip

\end{titlepage}


\section{Introduction}
\label{sec:intro}

The possible anomalous New Physics contributions to dipole and weak dipole moments
of the $\tau$ lepton bring renewed interest in development and revisiting charge-parity (CP)
violating signatures in the $\tau$-pair production in $Z$-boson decay at energies of the LHC.
Such process has been studied in the $pp$ collisions at the CERN LHC experiments in wide range of
invariant masses of outgoing $\tau$-lepton pair~\cite{ATLAS:2025oiy, CMS:2011ltz, ATLAS:2011ohj}.
What was not explored so far, is precise measurement of the transverse spin
correlations of the outgoing pair of $\tau$ leptons.

The {\tt TauSpinner} program~\cite{Czyczula:2012ny, Przedzinski:2018ett} is a convenient tool to study observables
sensitive to the New Physics (NP) effects in hadron colliders. It allows one to include NP and spin effects in case they
are absent in event samples generated with general purpose MC generators like {\tt Pythia}~\cite{Bierlich:2022pfr} 
or {\tt Sherpa}~\cite{Sherpa:2019gpd}. Program development has a long history driven by expanding
scope of its initially designed applications~\cite{Czyczula:2012ny}. 
First,  the longitudinal spin effects in case of  the Drell-Yan ($Z, W$) and the Higgs-decay processes 
of $\tau$ leptons at the
LHC~\cite{Czyczula:2012ny} were implemented.  Later, it was extended to applications for the NP interactions
in the hard processes, in which lepton pairs are accompanied with one or two hard jets~\cite{Kalinowski:2016qcd}, and to the transverse spin effects~\cite{Przedzinski:2014pla}.
The implementations was then further extended, Refs.~\cite{Richter-Was:2018lld,Richter-Was:2020jlt},
to allow study of  the electroweak effects and anomalous dipole moments of the $\tau$-lepton couplings
to vector bosons, Refs.~\cite{Banerjee:2023qjc, Korchin:2024appb}.

In recent years, anomalous dipole and week dipole 
moments~\cite{Beresford:2019gww,Haisch:2023upo,Beresford:2024dsc} of the $\tau$-lepton interaction with $\gamma$ and $Z$ boson gained
attention when searching for NP effects, because of indication of possible NP effects in case of the muon. 
In the paper~\cite{Korchin:2025vzx} we have presented an algorithm, 
implemented in {\tt TauSpinner} program~\cite{Richter-Was:2018lld},
to reweight previously generated events with the NP effects added in case of the 
$\gamma \gamma \to \tau \tau$ hard process.  These calculations  
include higher-order terms of dipole moments and phenomenological results. 
Now we return to  $\bar q q \to Z/\gamma^* \to \tau \tau$ process.
In the present paper, our focus is on discussing similar algorithms applied in the region around $Z$-boson peak
in the $\bar q q \to \tau \tau$ process. We evaluate numerical effects due to NP on elements of spin-correlation
matrix and show example of experimentally accessible distributions which are sensitive to spin correlations.

We first recall, Section~\ref{sec:theory}, details of formalism used to determine spin amplitudes and remind how
they lead to defining spin-correlation matrix, which being contracted with polarimetric vectors of the $\tau$ decays 
can be used for introducing spin correlations of $\tau$-decay products.
The expressions for spin amplitudes in case of the Standard Model (SM) in Born Approximation (BA) 
and NP extensions with $\tau$-lepton dipole moments (in coupling to photon) and weak dipole moments 
(in coupling to $Z$ boson) were given in ~\cite{Banerjee:2023qjc}.
Now we complete them with explicit  formulas of Improved Born Approximation (IBA), which includes electroweak (EW) 
corrections as in the formalism outlined in ~\cite{Bardin:1999yd}. Inspired by the measurements and publications of results of  the LEP experiments~\cite{ALEPH:1997wux, DELPHI:1997ssw},
we also introduce phase-shift between vector and axial-vector $Z \tau \tau$ couplings in the formalism of IBA, 
as a possible consequence of NP effects, and evaluate impact on the amplitudes and spin correlations.

The spin correlations and spin-correlation matrix $R_{ij}$ are first discussed in Subsection~\ref{subsec:qbar q},  
extension of $R_{ij}$ with the EW corrections is presented in Subsection ~\ref{sec:SM_Rij}, the
phase-shift between vector and axial-vector couplings is discussed in Subsection~\ref{subsec:SU2}, 
and some discussion on related measurement by the LEP experiments is given in Appendix~\ref{app:ALEPH}. 
Algorithm for event reweighting  in {\tt TauSpinner} is presented in Section \ref{sec:algorithm}.
Subsection~\ref{subsec:conv} covers issues of convention matching between the amplitudes 
presented in Subsection~\ref{sec:SM_Rij} and implementation
in {\tt TauSpinner} program, and configurations with high $p_T$ jets are briefly addressed in Subsection~\ref{subsec:jets}. Numerical results for spin-correlation matrix
are given in Subsection~\ref{sec:Rij} and impact of NP effects on a few interesting experimental distributions
is discussed in Subsection~\ref{sec:rhorho}. Section~\ref{sec:Outlook} closes the paper.
In Appendix~\ref{app:TauSpinner} we discuss technical details how to configure initialisation of {\tt TauSpinner}
with discussed NP effects in weights calculations.

\section{Amplitudes and spin correlations}
\label{sec:theory}

The formalism for calculating spin amplitude and spin correlations in $\bar q q \to Z/\gamma^* \to \tau \tau$
process, as implemented in the {\tt TauSpinner} algorithms,
was discussed in~\cite{Banerjee:2022sgf, Banerjee:2023qjc} for the SM  and the SM extensions including
dipole and weak dipole moments in the couplings $\gamma \tau \tau$ and $Z \tau \tau$.
The formalism for introducing to {\tt TauSpinner} algorithms of IBA and EW corrections
encapsulated into form-factors was discussed  in~\cite{Richter-Was:2018lld}.
Let us now consolidate this formalism and discuss how the existing implementations can be used to predict
impact of NP effects on spin correlations in  $\bar q q \to Z/\gamma^* \to \tau \tau$ process.
We do not define specifically what are the NP models, we just consider the effects of
form-factors which they can bring:
\begin{itemize}
\item
modification of the $\gamma \tau \tau$ vertex with the structure defined as for anomalous magnetic and/or electric form-factors;
 \item 
modification of the $Z \tau \tau$ vertex with the structure defined as for anomalous weak magnetic and/or weak electric form-factors;
\item
introduction of a phase-shift between vector and axial-vector couplings in the $Z \tau \tau$ vertex.
\end{itemize}

We will briefly summarise the formalism used in the following subsections.
 
 
\subsection{\texorpdfstring{Matrix element in $\bar{q}\, q \to Z/\gamma^* \to \tau \tau$ processes}{}}
\label{subsec:qbar q}

In this subsection details of spin correlations in the process $ q(k_1) + \bar{q}  (k_2)  \to 
\tau^-  (p_-) + \tau^+ (p_+)$  are presented, where $q$ stands for  the $u,  d,  s$ quarks.
We include $s$-channel exchange of  $Z$ boson,  virtual photon ($\gamma^*$) and their interference.

The $\gamma \tau \tau$ electromagnetic vertex is defined in the form 
\begin{equation}
\Gamma^\mu_\gamma (s)  = -i e \, Q_\tau \Big\{ \gamma^\mu F_1(s)  + \frac{\sigma^{\mu \nu} q_\nu}{2 m_\tau} \,  
\big[ i  A(s)   + B(s)  \gamma_5  \big] \Big\},
\label{eq:003}
\end{equation}
where   $e = \sqrt{4 \pi \alpha}$ is the elementary charge,  and $\alpha$ is the fine-structure constant,   
$Q_\tau=-1$, $\sigma^{\mu \nu}= \frac{i}{2} [\gamma^\mu, \, \gamma^\nu]$,  
$F_1 (s)$ is the Dirac form-factor, 
$A(s) =F_2(s)$ is the Pauli form-factor, and $B(s) = F_3(s)$ is the electric dipole form-factor, 
all of which depend on $s = q^2$, where $q=k_1+k_2=p_- + p_+$.  
At the real-photon point, $F_1(0)=1$, $A(0)$ is anomalous magnetic dipole moment $\mu_\tau$, 
while $B(0)$ is related to the $CP$-violating electric dipole moment $d_\tau$,  
\begin{equation}
A(0) = \frac{1}{2} (g_\tau-2) = \mu_\tau,    \qquad \quad B(0)  = \frac{2m_\tau}{e Q_\tau} d_\tau,
\label{eq:004}
\end{equation}
where $g_\tau$ is the gyro-magnetic factor. In (\ref{eq:003}) we choose $F_1(s) = 1$, and discuss EW 
radiative corrections below.    

To separate SM contribution from NP we explicitly include   
the QED contribution to $A(s)$ of the first order in $\alpha$~\cite{Berestetskii:1982qgu}
\begin{equation}
A(s)_{QED}=\frac{e^2 m_\tau^2}{4 \pi^2  s  \beta} \, \Big(  \log \frac{1-\beta}{1+\beta} + i \, \pi \Big).
\label{eq:005}
\end{equation}
Therefore we have  
\begin{equation}
A(s) = A(s)_{QED} + A(s)_{NP}, \quad \quad B(s) = B(s)_{NP}.
\label{eq:017}
\end{equation}  
We neglect in Eq.~(\ref{eq:017})  very small SM contribution to $B(s)$. Velocity of the $\tau$ lepton 
in the center-of-mass frame is denoted as $\beta = (1-4 {m_\tau^2}/{s})^{1/2}$.

The  $Z \tau \tau$ vertex is defined in the form  
\begin{equation}
\Gamma^\mu_Z (s)  =-i  \frac{g_Z}{2}  \, \Big\{ \gamma^\mu (v_\tau - \gamma_5 a_\tau) 
+ \frac{\sigma^{\mu \nu} q_\nu}{2 m_\tau} \,  \big[ i  X(s)  + Y(s)  \gamma_5  \big] \Big\},
\label{eq:018}
\end{equation}   
where $g_Z = e/(s_W c_W)$,  $s_W \equiv \sin \theta_W$, $c_W \equiv \cos \theta_W$ and $\theta_W$ is the weak mixing angle.  The vector and axial-vector coupling 
is respectively $v_\tau = T_{3 \tau} - 2 Q_\tau s^2_W$ and $a_\tau =T_{3 \tau}$, where  
$T_{3 \tau} = -1/2$ is the 3rd component of the weak isospin.    
Further, $X(s)$ is the weak anomalous form-factor, and $Y(s)$ is the $CP$ violating weak
electric form-factor.
On the $Z$-boson mass shell they are related to the weak anomalous magnetic, $\mu_\tau^{(w)}$, 
and weak electric, $d_\tau^{(w)}$, dipole moments, defined in ALEPH paper~\cite{ALEPH:2002kbp} via    
\begin{equation}
X(M_Z^2) = \mu_\tau^{(w)} (2 s_W c_W),  \qquad 
Y(M_Z^2) =  d_\tau^{(w)} (2 s_W c_W).  
\label{eq:ALEPH_moments}
\end{equation}
In the SM, $\mu_\tau^{(w)}$  was evaluated in Ref.~\cite{Bernabeu:1994wh}:   
 $\mu_{\tau, \, SM}^{(w)} = - (2.10 + i \, 0.61) \times 10^{-6}$.
It is therefore rather small, and this contribution is not explicitly given in our code, but can be 
included, if needed. The form-factors $X(s)$ and $Y(s)$ can be viewed as originating mainly from NP.   

We consider production of the polarised $\tau^+$ and $\tau^-$ leptons, which are characterized   
by the polarisation three-vectors in their rest frames, respectively 
\begin{equation}
s^+_j =   (s^+_1, \, s^+_2, \, s^+_3, \, 1 ),  \qquad  s^-_i = (s^-_1, \, s^-_2, \, s^-_3, \, 1 ),  
\label{eq:008}
\end{equation}
where $j, \,i = 1, 2, 3, 4$, and the Cartesian components $\vec{s}^{\,\pm} = (s^\pm_1, \, s^\pm_2, \, s^\pm_3)$ 
are defined with respect to the chosen reference frame\footnote{The frame is  
such that the axis ``3'' is along momentum of outgoing $\tau^-$, momentum of incoming quark $q$ lies in 
the plane ``1-3'',  and the axis ``2'' is orthogonal to the reaction plane. 
The following terminology is used in the following:  transverse (T) means component along axis 1,  normal (N) -- 
along axis 2, and longitudinal (L) -- along axis 3.}.  
For convenience we add in (\ref{eq:008}) the 4th components of the vectors, namely $1$.  

The matrix element squared and averaged over polarisations of the initial quarks   
\begin{equation}
|{\cal M}|^2 = |{\cal M}_\gamma|^2 + |{\cal M}_Z|^2 + 2 {\rm Re} ({\cal M}_\gamma^* \, {\cal M}_Z) 
\label{eq:022}
\end{equation}  
determines the differential cross section
\begin{equation}
\frac{d \sigma }{d \Omega} (q \, \bar q \to \tau^- \tau^+)= \frac{\beta}{64 \pi^2   s } 
|{\cal M}|^2,
\label{eq:023}
\end{equation}
where the mass of the quark is neglected.   

For the polarised $\tau$ leptons, each part of Eq.~(\ref{eq:022}) is expressed via polarisation 
vectors of $\tau$ leptons in their respective rest frames,  Eqs.~(\ref{eq:008}). Therefore
\begin{equation}
|{\cal M}|^2 = \sum_{i, j=1}^4 \, \bigl( R^{(\gamma)}_{i j} + R^{(Z)}_{i j} 
+ R^{(Z \gamma)}_{i j} \bigr) \, s^-_i  s^+_j ,
\label{eq:024}
\end{equation}
where $R^{(\gamma)}_{i j} $, $R^{(Z)}_{i j}$ and  $R^{(Z \gamma)}_{i j}$ are 
the matrices respectively of the $\gamma$ exchange, $Z$-boson exchange and $Z \gamma $ interference. 
In present implementation we include only terms linear in the form-factors $A(s), B(s), X(s)$ and $Y(s)$.
The expressions for $R_{ij}$ in BA can be found in~\cite{Banerjee:2023qjc}.

Next, we introduce EW corrections, following Refs.~\cite{Richter-Was:2018lld,Bardin:1999yd}. 
The amplitude in IBA, without dipole moments, can be written as 
\begin{eqnarray}
\label{eq:IBA}
{\cal M}^{IBA} &=& \frac{\, e^2}{s}  Q_q Q_\tau \, V_{\tau q} (s,t) \, \gamma_\mu \otimes \gamma^\mu  \nn \\
&& +
\Bigl(\frac{g_Z}{2}\Bigr)^2  \frac{Z_{\tau q}(s,t)}{d(s)} \, \gamma_\mu 
[v_q (s,t) - a_q \gamma_5] \otimes  \gamma^\mu [v_\tau (s,t) - a_\tau \gamma_5]  
\end{eqnarray} 
with the shortcut notation:  operator on the left of $\otimes$ is sandwiched between 
the spinors of the quarks, $ \bar{v}(k_2) \ldots u(k_1)$, and operator on the right of $\otimes$ is sandwiched 
between the spinors of the $\tau$ leptons, $\bar{u}(p_-) \ldots v(p_+)$.  
In (\ref{eq:IBA}) $Q_q$ is the charge of the quark $q$ in units of $e$.

The weak vector couplings $v_q(s,t)$ and $v_\tau(s,t)$ in IBA are complex and $s, \, t$ dependent, 
\begin{eqnarray}
v_q (s,t) &=& T_{3 q} - 2 Q_q s_W^2  K_q (s, t),  \nn \\
v_\tau (s,t) &=& T_{3 \tau} - 2 Q_\tau s_W^2 K_\tau (s, t),   \label{eq:vIBA}
\end{eqnarray}
while the axial-vector couplings are real and constant, $a_q = T_{3q}$ and $a_\tau = T_{3 \tau}$. 
The Mandelstam variable  $t = (k_1 - p_-)^2 = m_\tau^2 - s (1 - \beta \cos \theta)/2$, and it  
reduces to $ t \approx - {s} (1 - \cos \theta)/2 $ in the limit $m_\tau \approx 0$. 
For more detail we refer to original articles \cite{Richter-Was:2018lld, Richter-Was:2020jlt} 
and \cite{Banerjee:2023qjc}. 

Furthermore, $d(s) = s - M_Z^2 + i s \, \Gamma_Z / M_Z$ with running $Z$-boson decay 
width, and  $Z_{ \tau q} (s, t)$ is the normalisation correction for the $Z$-boson propagator
defined in Ref.~\cite{Richter-Was:2018lld}.

For technical reasons, of algebraic manipulation program, we can not add EW corrections
of $vv_{if}$ to $Z$ exchange amplitude, because it simultaneously depends on the incoming and 
outgoing fermion flavour, thus can not be absorbed into couplings redefinition.
Fortunately, as it affects vector $\times$ vector coupling only, part of the EW
corrections can be moved into an effective $\gamma$-exchange amplitude (the first term in (\ref{eq:IBA})).
Of course for that $Z$-boson (not $\gamma$) exchange correction, the photon propagator $1/s$ has to be replaced
with $1 / d(s)$, and the coupling $e^2$ with $(g_Z/2)^2$. We finally obtain for  the factor
$V_{\tau q} (s,t)$ in (\ref{eq:IBA})
\beq
V_{\tau q}(s,t) = \Gamma_{vp} +  \Bigl( \frac{g_Z}{e} \Bigr)^2 s_W^4  Z_{\tau q}(s,t) \frac{s}{d(s)}
\bigl[K_{\tau q}(s,t) - K_\tau (s,t) K_q (s,t) \bigr], 
\label{eq:Vtauq}
\eeq
which include $vv_{if}$ contribution from $Z$ on top of $\Gamma_{vp}$, where normalisation correction
$\Gamma_{vp}$ of photon exchange includes  re-summed vacuum-polarisation loop contribution.
The complex EW form-factors to vector couplings,  $K_q(s, t), \,  K_\tau(s, t)$ and $K_{\tau q}(s, t)$, are defined
in Ref.~\cite{Richter-Was:2018lld}. 

Eq.~(\ref{eq:IBA}) represents improved (with EW corrections) photon exchange amplitude with running QED constant 
and improved $Z$-boson exchange which include:
\begin{itemize}
\item[--] corrections to the photon propagator coming from the vacuum-polarisation loops,
\item[--] corrections to the $Z$-boson propagator and couplings embedded in the form-factors $Z_{\tau q}(s,t)$,
$K_q (s,t), \, K_\tau (s,t)$ and $K_{\tau q}(s,t)$, 
\item[--] contributions from the $WW$- and $ZZ$-box diagrams also included in the form-factors,
\item[--] mixed ${\cal O}(\alpha \alpha_s, \, \alpha \alpha_s^2, \ldots)$ corrections originating from gluon 
insertions in the self-energy loop diagrams.
\end{itemize}  
The EW form-factor corrections are available in {\tt Dizet} library \cite{Bardin:1999yd}
(see further details in Ref.~\cite{Richter-Was:2018lld}).

We can now include the dipole form-factors $A(s)$, $B(s)$, $X(s)$ and $Y(s)$ in the amplitude with EW corrections.  Using Gordon identities (see details in~\cite{Banerjee:2023qjc}), the amplitude takes the form   
\begin{eqnarray}
\label{eq:DM}
&& {\cal M}^{DM} = \frac{e^2 }{s} Q_q Q_\tau \,  V_{\tau q} (s,t) \, \gamma_\mu \otimes \bigl[ A \gamma^\mu  +
\frac{(p_+ - p_-)^\mu}{2m_\tau} (A - i B \gamma_5) \bigr]  \nn \\
&& +
\Bigl(\frac{g_Z}{2}\Bigr)^2  \frac{Z_{\tau q}(s,t)}{d(s)}  \, \gamma_\mu 
[v_q (s,t) - a_q \gamma_5] \otimes  \bigl[X \gamma^\mu  +\frac{(p_+ - p_-)^\mu}{2m_\tau} (X - i Y \gamma_5)
\bigr], 
\end{eqnarray}
and the total amplitude is $ {\cal M} = {\cal M}^{IBA} + {\cal M}^{DM}$.

Note, that EW corrections in IBA become numerically sizable for the cross-section at high energies, 
well above $Z$-boson peak.  
At the lower energies, if precision required is within a few percent, one can use amplitudes in which the EW  
form-factors  $Z_{\tau q}(s,t)$, $K_q(s,t)$,  $K_\tau (s,t)$,  $K_{\tau q}(s,t)$ as well as $\Gamma_{vp} (s)$ and 
$V_{\tau q} (s,t)$ are equal to 1, and the couplings $v_\tau, v_q, a_\tau, a_q, \alpha$ are 
the {\it effective} ones of the $Z$-pole (PDG values~\cite{PDG:2024cfk}).
See also discussion in Appendix~\ref{app:ALEPH}.  However, due to complicated structure of the transverse spin correlations, it might not longer be the case that the BA amplitude is sufficient, see Fig.~\ref{Fig:rxx_rxy_SM} 
in Section~\ref{sec:numerical}.

In the following, to simplify formulas,  we will not indicate explicitly $(s, t)$-dependence of  
$v_q (s,t)$, $v_\tau (s,t)$, $Z_{\tau q} (s,t)$ and $V_{\tau q} (s,t)$.

\subsection{\texorpdfstring{Spin-correlation matrix $R_{ij}$}{}}
\label{sec:SM_Rij}

The complete matrix element in the IBA, $ {\cal M} = {\cal M}^{IBA} + {\cal M}^{DM}$, has been used to derive, with  {\tt Mathematica} package, fortran code for the elements $R_{ij}$ of the 
spin-correlation matrix.
Here we would like to recall explicit  analytical form for a few selected elements $R_{ij}$ of particular interest.

The elements $R_{i j}$ are explicitly given in Ref.~\cite{Banerjee:2023qjc} in the BA and include anomalous dipole moments. Here we elaborate on using ${\cal M}^{IBA}$ (SM with EW corrections) for 
selected  $R_{ij}$ components.  In the considered kinematics around $Z$-boson pole it is sufficient 
to take the limit for the Lorentz factor $\gamma = E_\tau/m_\tau  = 
\sqrt{s}/{2 m_\tau} \gg 1$ and the $\tau$ velocity $\beta \approx 1$. We also introduce
modulo squared $Z$-boson propagator:  $D_Z(s) = |d(s)|^2 = (s-M_Z^2)^2 + s^2 \Gamma_Z^2/M_Z^2$.

In the formulas for $R_{ij}$ below we explicitly assume that $a_q, a_\tau$ are real, while $v_q, v_\tau$
can be complex. For the transverse-transverse (TT) (1-1), transverse-normal (TN) (1-2), 
and normal-normal (NN) (2-2) spin-correlation elements we obtain:
\bea
\label{eq:R11}
R_{11}^{(\gamma), \, IBA} &=&  \frac{e^4}{4} \, Q_q^2 Q_\tau^2 \, |V_{\tau q}|^2 
 \sin^2 \theta,  \nn \\ 
R_{11}^{(Z), \, IBA} &=&   \frac{e^4 \, s^2  }{64 s_W^4 c_W^4 {D}_Z(s)} |Z_{\tau q}|^2
(a_q^2 + |v_q|^2) (|v_\tau|^2 - a_\tau^2)  \sin^2 \theta,  \nn  \\  
R_{11}^{(Z \gamma), \, IBA}  &=&  \frac{e^4 \,  s }{8 s_W^2 c_W^2 {D}_Z (s)} 
\,  Q_q Q_\tau \, \nn \\
&& \times \Bigl[ (s-M_Z^2) {\rm Re} \bigl(v_\tau v_q \, V^*_{\tau q} \, Z_{\tau q} \,  \bigr) 
+ s  \frac{\Gamma_Z}{M_Z} {\rm Im}  \bigl(v_\tau v_q \, V^*_{\tau q} \, Z_{\tau q} \bigr) \Bigr]
 \sin^2 \theta ,    
\eea

\bea
\label{eq:R12}
R_{12}^{(\gamma), \, IBA} &=& 0, \nn \\
R_{12}^{(Z), \, IBA} &=&  - \frac{e^4 \, s^2  }{32 s_W^4 c_W^4 {D}_Z(s)} |Z_{\tau q}|^2 \,
(a_q^2 + |v_q|^2) \, a_\tau \, {\rm Im}(v_\tau)  \sin^2 \theta,  \nn \\ 
R_{12}^{(Z \gamma), \, IBA} &=& \frac{e^4 \,  s }{8 s_W^2 c_W^2 {D}_Z (s)} 
 \, Q_q \, Q_\tau \,   \nn \\   
&& \times a_\tau  \Bigl[ (s-M_Z^2)  {\rm Im}  \bigl(v_q \, V^*_{\tau q} \, Z_{\tau q} \bigr)   
 - s  \frac{\Gamma_Z}{M_Z} {\rm Re}  \bigl(v_q \, V^*_{\tau q} \, Z_{\tau q} \bigr)
 \Bigr]   \sin^2 \theta. 
\eea

For the longitudinal polarisation elements one finds
\bea
\label{eq:R34}
R_{34}^{(\gamma), \, IBA} &=& 0, \nn \\
R_{34}^{(Z), \, IBA} &=&  - \frac{e^4 \, s^2  }{32 s_W^4 c_W^4 {D}_Z(s)}  |Z_{\tau q}|^2 \,  \nn \\
&& \times \Bigl[ a_\tau (a_q^2+|v_q|^2) {\rm Re}(v_\tau)  (1+\cos^2 \theta)   
 + 2 a_q (a_\tau^2 + |v_\tau|^2) {\rm Re}(v_q) \cos \theta \Bigr],   
\nn \\
R_{34}^{(Z \gamma), \, IBA} &=& 
- \frac{e^4 \,  s }{8 s_W^2 c_W^2 {D}_Z (s)}  \, Q_q Q_\tau  \nn \\
&&\times \Bigl\{ a_\tau \Bigl[ (s-M_Z^2) {\rm Re}(v_q \, V_{\tau q}^* \, Z_{\tau q})   
 + s \frac{\Gamma_Z}{M_Z} {\rm Im}(v_q \, V_{\tau q}^* \, Z_{\tau q}) \Bigr]  (1+ \cos^2 \theta) \nn \\  
 && + 2 a_q   \Bigl[ (s-M_Z^2) {\rm Re}(v_\tau \, V_{\tau q}^* \, Z_{\tau q}) 
 + s \frac{\Gamma_Z}{M_Z} {\rm Im}(v_\tau \, V_{\tau q}^* \, Z_{\tau q}) \Bigr] \cos \theta \Bigr\}. 
\eea

Finally, the elements $R_{44}$ are
\bea
\label{eq:R44}
R_{44}^{(\gamma), \, IBA} &=&  \frac{e^4}{4} Q_q^2 Q_\tau^2  |V_{\tau q}|^2 (1+ \cos^2 \theta), \nn \\
R_{44}^{(Z), \, IBA} &=&  \frac{e^4 \, s^2  }{64 s_W^4 c_W^4 {D}_Z(s)} |Z_{\tau q} |^2 \nn \\
&& \times \Bigl[ (a_q^2 + |v_q|^2) (a_\tau^2 + |v_\tau|^2) (1+ \cos^2 \theta)   
+ 8 \, a_q  a_\tau  {\rm Re}(v_q) \, {\rm Re}(v_\tau)  \cos \theta \Bigr],  \nn \\
 R_{44}^{(Z \gamma), \, IBA} &=&  \frac{e^4 \,  s }{8 s_W^2 c_W^2 {D}_Z (s)} 
Q_q Q_\tau \nn \\
&& \times \Bigl\{  \Bigl[ (s-M_Z^2)  {\rm Re}(V^*_{\tau q} \, Z_{\tau q} \, v_q v_\tau ) 
 + s  \frac{\Gamma_Z}{M_Z} {\rm Im } (V^*_{\tau q} \, Z_{\tau q} \, v_q v_\tau ) \Bigr] (1+\cos^2 \theta) \nn \\
&& + 2  \,  a_q a_\tau   \Bigl[(s-M_Z^2)  {\rm Re}(V^*_{\tau q} \, Z_{\tau q} ) 
 + s  \frac{\Gamma_Z}{M_Z} {\rm Im } (V^*_{\tau q} \, Z_{\tau q} )   \Bigr] \cos \theta \Bigr\}.      
\eea
In these equations $\theta$ is the scattering angle of outgoing $\tau^-$ lepton with respect to the quark $q$. 

In the IBA, the following symmetry relations hold near $Z$-boson peak
\begin{equation}
\label{eq:symmetry}
R_{22}^{IBA} = - R_{11}^{IBA},  \qquad R_{21}^{IBA} = R_{12}^{IBA}, \qquad 
R_{43}^{IBA} = R_{34}^{IBA}   
\end{equation} 
for the contributions from $\gamma$, $Z$ and $Z \gamma$.  

Note that to account for the quark color average, one should include additional factor 1/3 
in Eqs.~(\ref{eq:R11})-(\ref{eq:R44}). 
    
Omitting EW corrections means setting  
$\Gamma_{vp} = 1$, $Z_{\tau q}=K_{\tau q}=K_{\tau}=K_{q}=1$ and $V_{\tau q}=1$.
In this approximation  $v_q,  v_\tau$ are real and constant, and formally ${\cal M}^{IBA} =  {\cal M}^{BA}$,
i.e. matrix element in the BA.
If the {\it effective} couplings are used in the  ${\cal M}^{BA}$ calculation, most of the effects on the cross-section and longitudinal polarisation are restored. But not necessarily for TN correlation, which is sensitive 
to ${\rm Im}(v_\tau)$ in the leading term at the $Z$ pole.

\subsection{Introducing phase-shift between vector and axial-vector couplings}
\label{subsec:SU2}

The measurement of ALEPH Collaboration~\cite{ALEPH:1997wux} of TN and TT spin correlations allowed the authors 
to quantify limits on the phase-shift between vector and axial-vector $Z \tau \tau$ couplings.
Motivated by this measurement we included in the initialisation of the {\tt TauSpinner} program an option
which allows one to introduce a similar phase-shift.

The structure of ${\cal M}^{IBA}$ discussed in Section~\ref{subsec:qbar q} has been derived with convention for 
the EW form-factors being complex, and the lowest-order couplings being real.
Staying within this convention, we introduce the phase-shift $\Phi$ not directly to the definition of the
couplings themselves but as rescaling of the EW form-factors  in the following manner: 
\begin{eqnarray}
  \label{eq:Phi_tau_shift}
  && Z_{\tau q} \to Z^{\prime}_{\tau q }       =  Z_{\tau q}    \times e^{  - i \Phi},\nn  \\
  && K_\tau \to K^{\prime}_\tau   =  K_\tau    \times e^{  i \Phi}, \nn \\
  && K_{\tau q}  \to K^{\prime}_{\tau q}  =  K_{\tau q}  \times e^{  i \Phi}.  
\end{eqnarray}
The expression $(\ref{eq:IBA})$ of ${\cal M}^{IBA}$ can be now written as 
\begin{eqnarray}
\label{eq:IBA_phi}
      {\cal M}^{IBA} &\rightarrow& \frac{e^2}{s}  Q_q Q_\tau \, V_{\tau q} (s,t) \, \gamma_\mu \otimes \gamma^\mu  \nonumber \\ 
      && + \Bigl( \frac{g_Z }{2 } \Bigr)^2  \frac{Z_{\tau q}(s,t) }{d(s)} \, e^{ -i \Phi} \, \gamma_\mu
     [T_{3q}- 2 Q_{q} s_W^2 K_{q}(s, t) - a_q \gamma_5]  \nn \\
     && \, \otimes  \, \gamma^\mu \bigl[ T_{3\tau} - 2 Q_{\tau} s_W^2 K_{\tau}(s,t) e^{  i \Phi} - a_\tau \gamma_5
		\bigr]. 
 \end{eqnarray}
Note that factor $V_{\tau q}(s,t)$ in Eq.~(\ref{eq:Vtauq}) does not change under the transformation 
$(\ref{eq:Phi_tau_shift})$.
There is no need to re-derive expressions for $R_{ij}$ starting from the definition of ${\cal M}^{IBA}$
in (\ref{eq:IBA_phi}), instead we can redefine vector coupling for the $\tau$ lepton as follows 
\begin{equation}
 \label{eq:v_tau_prime}
 v_{\tau}      \rightarrow  v_{\tau}^{\prime} = T_{3\tau} - 2 Q_{\tau} s_W^2 K_{\tau}(s, t) e^{  i \Phi}. 
\end{equation}

Now we can express  elements $R_{ij}$, detailed in Section~\ref{sec:SM_Rij} (Eqs.~(\ref{eq:R11})-(\ref{eq:R44})), using $v_{\tau}^{\prime}$ instead of $v_{\tau}$. Note also that  
$| Z_{\tau q}(s,t)| =  | Z_{\tau q}(s,t) e^{-i \Phi}|$.
Using (\ref{eq:v_tau_prime}) we  can calculate the phase $\Phi_{v_\tau}$ of the vector coupling, defined via
$v_\tau^{\prime} = |v_\tau^{\prime}| e^{i \Phi_{v_\tau}}$,  and use equations
\begin{eqnarray}
\label{eq:v_tau_prime_expand}
{\rm Re} (v_\tau^{\prime} ) & = & T_{3 \tau} - 2 Q_{\tau} s_W^2 {\rm Re}(K_\tau(s,t)) \cos(\Phi) 
+  2 Q_{\tau} s_W^2  {\rm Im}(K_\tau(s,t))\sin(\Phi), \nn \\ 
{\rm Im}(v_\tau^{\prime}) & = &  - 2 Q_\tau s_W^2 {\rm Re}(K_\tau(s,t)) \sin (\Phi) 
- 2 Q_{\tau} s_W^2 {\rm Im}(K_\tau(s,t)) \cos(\Phi),  \nn \\
|v_\tau^{\prime}| & = & \bigl[{\rm Re}(v_\tau^{\prime})^2 + {\rm Im}(v_\tau^{\prime})^2 \bigr]^{1/2}
\end{eqnarray}
and the formulas  
\begin{equation}
\label{eq:Phi_tau_sin_cos}
\sin (\Phi_{v_\tau}) =  {\rm Im}(v_\tau^{\prime})/|v_\tau^{\prime}|,   \qquad \quad 
\cos (\Phi_{v_\tau}) =  {\rm Re}(v_\tau^{\prime})/|v_\tau^{\prime}|. 
\end{equation}

The phase $\Phi_{v_\tau}$ of the vector coupling comes from the complex form-factor $K_\tau(s,t)$ and the
phase-shift $\Phi$ imposed with the transformation~(\ref{eq:Phi_tau_shift}). The relation between  
$\Phi$ and $\Phi_{v_\tau}$ is not trivial but can be calculated analytically from (\ref{eq:v_tau_prime_expand}) 
and (\ref{eq:Phi_tau_sin_cos}), and then numerically using tabulated values of $K_\tau(s,t)$. 

In the same manner as Eq.~(\ref{eq:Phi_tau_shift}), one could introduce phase-shift in the 
$Z$-boson couplings to quarks, including the phase $\Phi$ into  $K_q(s,t)$ instead of $K_\tau(s,t)$:
\begin{eqnarray}
  \label{eq:Phi_q_shift}
  && Z_{\tau q} \to Z^{\prime}_{\tau q }       =  Z_{\tau q}    \times e^{ - i \Phi}, \nn \\
  && K_q \to K^{\prime}_q   =  K_q    \times e^{   i \Phi},  \nn \\
  && K_{\tau q}  \to K^{\prime}_{\tau q}  =  K_{\tau q}  \times e^{  i \Phi}.     
\end{eqnarray}

Correspondingly,  we can use expressions for $R_{ij}$ starting from definition of ${\cal M}^{IBA}$ in
 (\ref{eq:IBA_phi}), with the replacement
\begin{equation}
 \label{v_q_prime}
 v_q      \rightarrow  v_q^{\prime} = T_{3q} - 2 Q_{\tau} s_W^2 K_q(s, t)  e^{  i \Phi}. 
\end{equation}

The rest of the discussion will follow exactly the same path.
In this case the TT element $R_{11}^{IBA}$   
is not sensitive to the  phase-shift,  while  the TN element 
$R_{12}^{IBA}$ strongly depends on it (see Section \ref{sec:numerical}). 

Let us make an observation that in Eq.~(\ref{eq:IBA_phi}) the factor $e^{-i \Phi}$ 
multiplying $ Z_{\tau q}(s,t)$ can be absorbed into definition of the couplings, 
leading to the expression for ${\cal M}^{IBA}$: 
\begin{eqnarray}
\label{eq:IBA_phi_bis}
      {\cal M}^{IBA} &\rightarrow& \frac{e^2}{s}  Q_q Q_\tau \, V_{\tau q} (s,t) \, \gamma_\mu \otimes \gamma^\mu  \nn \\ 
      && + \Bigl( \frac{g_Z}{2} \Bigr)^2 \frac{Z_{\tau q}(s,t)}{d(s)}\, \gamma_\mu
     [T_{3q}- 2 Q_{q} s_W^2 K_{q}(s, t) - a_q \gamma_5]  \nn \\
     && \, \otimes  \, \gamma^\mu \bigl[ T_{3\tau} e^{  - i \Phi} - 2 Q_{\tau} s_W^2 K_{\tau}(s,t) - a_\tau  e^{ - i \Phi} \gamma_5 \bigr], 
\end{eqnarray}
which suggests redefining the axial-vector and vector couplings in the following way 
\begin{equation}
 \label{eq:v_tau_a_prime-prime}
 a_{\tau}  \rightarrow  a_{\tau}^{\prime\prime} =  a_{\tau}e^{  - i \Phi}, \quad   
 v_{\tau}   \rightarrow  v_{\tau}^{\prime\prime} = T_{3\tau} e^{  - i \Phi} - 2 Q_{\tau} s_W^2 K_{\tau}(s, t). 
\end{equation}
Eqs.~(\ref{eq:IBA_phi}) and  (\ref{eq:IBA_phi_bis}) are equivalent and result in the same phase difference
between vector and axial-vector couplings 
\begin{equation}
\label{eq:phase_difference}
  \Phi_{v_\tau^{\prime\prime}} - \Phi_{a_\tau^{\prime\prime}} =  \Phi_{v_\tau^{\prime}} - \Phi_{a_\tau^{\prime}}.
\end{equation}

\begin{figure}
  \begin{center}                               
{
   \includegraphics[width=7.2cm,angle=0]{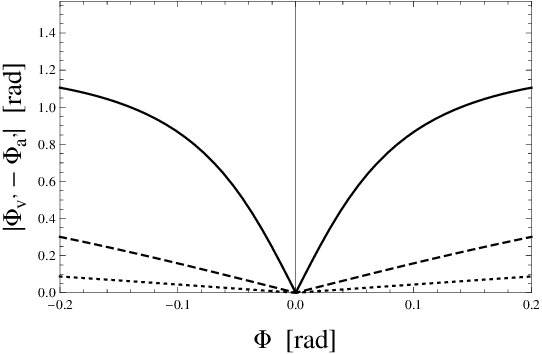}
}
\end{center}
\caption{Relation between  absolute value of phase difference $|\Phi_{v^\prime} -  \Phi_{a^\prime}|$ 
and phase-shift  $\Phi$ of transformation  (\ref{eq:Phi_tau_shift})
for $\tau$ lepton (solid line), and of transformation (\ref{eq:Phi_q_shift}) for up quark (dashed line) 
and down quark (dotted line).}
 \label{Fig:Phi_Phitau} 
\end{figure}

As an illustration, in Fig.~\ref{Fig:Phi_Phitau} we present $\Phi$-dependence of the absolute value of phase difference $|\Phi_{v'} -  \Phi_{a'}|$ for $\tau$ lepton,  $u$  and $d$ quarks. 
It was calculated from (\ref{eq:v_tau_prime_expand}) and  (\ref{eq:Phi_tau_sin_cos}) 
for the $\tau$ lepton, and (\ref{v_q_prime}) for the quarks. In this figure we set   
$K_\tau (s,t) = K_q (s,t) = 1.0$ and $s^2_W =  0.23151$. It is seen that $\Phi$-dependence 
of $|\Phi_{v'} -  \Phi_{a'}|$ for $\tau$ lepton is more pronounced than that for quarks.

\section{The reweighting algorithm for {\tt TauSpinner} }
\label{sec:algorithm}

\subsection{\texorpdfstring{Adjusting convention of  $|{\cal M}|^2$ calculations and TauSpinner}{}}
\label{subsec:conv}                                                                                                                                 

One can further rearrange Eqs.~(\ref{eq:023}), (\ref{eq:024}) and introduce normalised elements 
$r_{ij} = R_{ij}/R_{44}$,
factorising out explicitly spin correlations component of the total cross-section, as shown below
\begin{equation} 
\frac{d \sigma}{d \Omega} (\bar q q   \to \tau^- \tau^+) = \frac{\beta}{64 \pi^2   s } 
\, R_{44} \; \Bigl( r_{tt} +\sum_{i, j=1}^3 \, r_{i j}  \, s^-_i  s^+_j  \Bigr), \qquad r_{tt}=1,
\label{eq:016}
\end{equation}
where $\beta = (1-4 m_\tau^2/s)^{1/2}$.

Let us stress, that the frame and sign convention of $R_{ij}$ presented in ~\cite{Banerjee:2023qjc} and formulas
Eqs.~(\ref{eq:R11})-(\ref{eq:R44}) differ from the ones used in the {\tt TauSpinner} program,
when contracted with the $\tau$ leptons polarimetric vectors (recall Eq.~(\ref{eq:024})).
The change in convention which is performed internally in the code reads as follows:  
\begin{eqnarray}
\label{eq:framesR}
 R_{tt}\;\leftarrow \;\; R_{44}, & R_{tx}\leftarrow-R_{42}, & R_{ty}\leftarrow -R_{41},\;  R_{tz}\leftarrow-R_{43}, \nonumber \\
 R_{xt}\leftarrow -R_{24}, & R_{xx}\leftarrow  \;\;R_{22}, & R_{xy}\leftarrow  \;\;  R_{21},\;  R_{xz}\leftarrow \;\; R_{23}, \nn \\
 R_{yt}\leftarrow -R_{14}, & R_{yx}\leftarrow \;\;  R_{12}, & R_{yy}\leftarrow  \;\;  R_{11},\;  R_{yz}\leftarrow \;\; R_{13},  \nonumber \\
 R_{zt}\leftarrow -R_{34}, & R_{zx}\leftarrow \;\; R_{32}, & R_{zy}\leftarrow  \;\; R_{31}, \;  R_{zz}\leftarrow \;\; R_{33}.
  \end{eqnarray}

There are several reasons for that frame orientation differences  used in {\tt TauSpinner}
and {\tt Tauola} decay library~\cite{Jadach:1993hs} as well. Historical
one; in the past, reactions were organized having $z$ axis along $e^+$ or antiquark direction, whereas
now, many authors prefer to use  $z$ axis along $e^-$ or quark direction. This past choice was  used
for {\tt KKMC}~\cite{Arbuzov:2020coe} and {\tt Tauola}~\cite{Jadach:1993hs} programs. To adjust, this require
$\pi$ angle rotation around axis usually perpendicular to the reaction plane. There is also an overall
sign which is affecting $\tau^+$ spin indices of the spin-correlation matrices. This is moved in
all {\tt Tauola/TauSpinner} interfaces~\cite{Davidson:2010rw}  from $\tau$ decay to spin-correlation matrices.
Also, for many past calculations which we rely on as a reference, rest
frames of $\tau^\pm$   were chosen to have the common $z$ axis direction, parallel to $z$ axis of
reaction frame (where incoming partons are not defining $z$ direction). For calculations involving
NP, we have found, that for present day authors it is often convenient to allow for distinct frame orientations
than that. The easiest way to solve this different conventions was to provide for {\tt TauSpinner} implementation
an adjustment internal routine.

There is also another adjustment, this time  for $\tau^+$ polarimetric vector orientation, also of
its overall sign. At present,
this adjustment is  shifted into $R_{ij} $ matrix redefinition too\footnote{It is done separately,
in different place of the code, just before the event weight calculation.}, even though it does not correspond to change of
its orientation, but is for  $\tau^+$ polarimetric vector. Finally
\begin{eqnarray}
\label{eq:frames}
 R_{tt}= \;\; R_{44}, & R_{tx}=     -R_{42}, & R_{ty}=     -R_{41},\;  R_{tz}=     -R_{43}, \nonumber \\
 R_{xt}=     -R_{24}, & R_{xx}= \;\; R_{22}, & R_{xy}= \;\; R_{21},\;  R_{xz}= \;\; R_{23}, \nn          \\
 R_{yt}= \;\; R_{14}, & R_{yx}=     -R_{12}, & R_{yy}=     -R_{11},\;  R_{yz}=     -R_{13}, \nonumber \\
 R_{zt}=     -R_{34}, & R_{zx}= \;\; R_{32}, & R_{zy}= \;\; R_{31},\;  R_{zz}= \;\; R_{33}.  
  \end{eqnarray}

With this transformation\footnote{That means that TT, NN and TN correspond, respectively, to 
$R_{yy}$, $R_{xx} $ and and $-R_{yx}$.}, 
expressions for the $R_{ij} $ elements of Eq.~(\ref{eq:frames}) with $i, j=t,x,y,z$ 
are used  in {\tt TauSpinner} event  reweighting algorithm discussed in Section~\ref{sec:algorithm}
for calculating weight implemented in {\tt TauSpinner} code and in Section~\ref{sec:numerical} for presenting numerical results.

The basic formalism  of {\tt TauSpinner} is documented in Ref.~\cite{Przedzinski:2018ett}, 
Section 2.2,  Eqs.~(7) to (12). We do not repeat details of this formalism here, nor details how kinematics
of hard process is deciphered from the kinematics of the $\tau$-decay products. We recall however
a few basic equations for calculating final weights, which allow to take into account changes in the cross-section
and spin correlations in the SM and SM+NP models.

The  basic  equation in the calculation of the cross-section is
 \begin{eqnarray}
&d \sigma = \sum_{flav.} \int dx_1 \, dx_2 \, f(x_1,...)\, f(x_2,...) \, d\Omega^{parton\; level}_{prod} \; d\Omega_{\tau^+} \; d\Omega_{\tau^-} \nonumber \\
& \times \Bigl(\sum_{\lambda_1,  \lambda_2 }|{\cal M}^{prod}_{parton\; level}|^2 \Bigr)
 \Bigl(\sum_{\lambda_1 }|{\cal M}^{\tau^+}|^2 \Bigr)
 \Bigl(\sum_{\lambda_2 }|{\cal M}^{\tau^-}|^2 \Bigr) \, wt_{spin}, 
\label{eq:parton-level}
\end{eqnarray}
 where $x_1$, $x_2$ denote fractions of the beam momenta carried by the partons, $f(x_1, \ldots)$, $f(x_2, \ldots)$
 denote parton density functions (PDF)s of the beams\footnote{The {\tt TauSpinner} algorithm does not use information of the
 flavour of incoming partons from the event record, allowing for application of its weight also on experimental data.}
 and $d\Omega$ denote phase-space integration elements respectively for $2 \to 2$ hard process and $\tau$ decays.
 Eq.~(\ref{eq:parton-level}) represents product of distributions for the $\tau^\pm$ production and decay: 
$ \Bigl(\sum_{\lambda_{1,2} }|{\cal M}^{\tau^\pm}|^2 \Bigr)$
 stands for the decay matrix element squared, and 
$ \Bigl(\sum_{\lambda_1,  \lambda_2 }|{\cal M}^{prod}_{parton\; level}|^2 \Bigr)$  for production matrix element squared. Only the  spin weight $wt_{spin}$ needs input both from $\tau^\pm$ production and decay.

Elements $R_{i j}$ used in calculation of components of Eq.~(\ref{eq:parton-level}) are taken as weighted average
(with PDFs and production matrix elements squared) 
over all flavour configurations, as in the following equation:
{\small
 \begin{eqnarray}
R_{i j} \to \frac{ \sum_{flav.} f(x_1, \ldots) f(x_2, \ldots) 
\Bigl(\sum_{\lambda_1,  \lambda_2 }|{\cal M}^{prod}_{parton\ level}|^2 \Bigr)  R_{i j} }
{\sum_{flav.}  f(x_1, \ldots)f(x_2, \ldots) 
\Bigl(\sum_{\lambda_1,  \lambda_2 }|{\cal M}^{prod}_{parton\ level}|^2 \Bigr)\;\;\;\;\;\;\;} . \label{eq:Rij-ave}
 \end{eqnarray}
}
No new approximation is introduced in this  way, the denominator of Eq.~(\ref{eq:Rij-ave}) cancels explicitly the corresponding factor of Eq.~(\ref{eq:parton-level}).

For the $wt_{spin}$ calculation, the normalised  elements $ r_{i j} = R_{i j}/R_{tt}$ are used, 
following Eq.~(\ref{eq:016})
\begin{equation}
wt_{spin} = \sum_{i ,j=t,x,y,z} r_{i j} h^i_{\tau^+} h^j_{\tau^-}. \label{eq:wtspin}
\end{equation}
 Here $ h^i_{\tau^+}, \,  h^j_{\tau^-}$ stand for decay-mode dependent $\tau$ polarimetric vectors.
Note, that $wt_{spin}$ is independent of the PDFs, except through already
averaged over partons contribution from elements $ r_{i j}$ of spin correlations matrix.
 
To introduce the corrections due to different spin effects and modified production
process in the generated sample (i.e. without re-generation
of events),  one can define the weight $wt$, representing
the ratio of the new to old cross-sections at each point in the phase space.

Eq.~(\ref{eq:parton-level}) for the modified cross-section takes then the form
 \begin{equation}
   d \sigma_{new} = d\sigma_{old} \; wt_{prod}^{new/old} \; wt_{spin}^{new/old} , 
	\label{eq:sigma_new}
 \end{equation}
 where $ d\sigma_{old}$ is calculated with Eq.~(\ref{eq:parton-level}),  $wt_{spin}$ using Eq.~(\ref{eq:wtspin}) and 
$ wt_{prod}^{new/old}$ using Eq.~(\ref{eq:wtprod}) below
{\small
\begin{equation}
wt_{prod}^{new/old}=  \frac{\sum_{flav.}f(x_1, \ldots) f(x_2, \ldots) \Bigl(\sum_{spin }|{\cal M}^{prod}_{part. lev.}|^2\Bigr) \Big|_{new}}
{\sum_{flav.}f(x_1, \ldots) f(x_2, \ldots) \Bigl(\sum_{spin }|{\cal M}^{prod}_{part. lev.}|^2\Bigr) \Big|_{old}}\\
= \frac{R_{tt}|_{new}}{R_{tt}|_{old}}.
\label{eq:wtprod}
\end{equation}
Present implementation  assumes, that the generated sample has no spin correlations included, however it can
easily be extended\footnote{Such special case is available for $\bar q q \to \tau \tau$ processes, where, e.g.
polarisation but not spin correlations, was included in the generated sample.}
to provide weight calculated as
\begin{equation}
wt_{spin}^{new/old} =\frac{ \sum_{i ,j=t,x,y,z} r_{i j} h^i_{\tau^+} h^j_{\tau^-}\Big|_{new}}{ \sum_{i ,j=t,x,y,z} r_{i j} h^i_{\tau^+} h^j_{\tau^-}\Big|_{old}} . \label{eq:wtspinr}
\end{equation}

The  {\tt TauSpinner} program provides both weights, $ wt_{spin}$ and $wt_{prod}$,  which allows to modify
per-event distributions of sample generated according to $ d\sigma_{old}$ model. 
The  combined weight should be used as multiplicative product 
\begin{equation}
wt = {wt_{prod}^{new/old}} \; \times \; {wt_{spin}^{new/old}}, \label{eq:combi}
\end{equation}
where the first term of the weight represent modification of the matrix elements for production,
the second one -- of the spin correlations. It is nothing else than ratios of spin averaged amplitudes squared 
for the whole process; new to old. If the analysis is sensitive to changes in the PDF parametrisations
used for sample generation and {\tt TauSpinner} weights calculations, it should be taken
into account in calculation of $wt_{prod}^{new/old}$ and  $wt_{spin}^{new/old}$ in Eq.~(\ref{eq:combi}).
In case production process is not modified, $wt_{prod}^{new/old}$ is equal to 1.
In case of originally generated sample without spin correlations, $wt_{spin}^{new/old}$ alone allows one 
to introduce the desired spin effects. 

\subsection{\texorpdfstring{Configurations with high $p_T$ jets}{}}
\label{subsec:jets}

Our matrix elements and tests observables discussed later rely on
simplified kinematics. This would lead to no ambiguities if only parton-level
$2 \to 2 $ processes would be considered. For the general
case, this requires attention. In Ref.~\cite{Richter-Was:2016mal}, the question of
separating hard (EW process) was addressed, and it was found that with
a proper choice of reference frames one could get the impact of hadronic jets
activity absorbed into adjustment of reference frames. This is possible because
of properties of QED and QCD matrix elements and it was explored in
preparation for the measurements of the $Z$ and the $W^\pm$ properties with the help of
templates techniques~\cite{Richter-Was:2016avq}. The properties seem to be quite 
general and originating from the properties of the Lorentz group and its representations.

In present applications, we rely on the transverse degrees of freedom, and
underlying to our methods separation out of initial-state hadronic (jets)
activity needs to attract attention. Especially if events of high $p_T$ jets
are present in the analyzed samples. Then, our two choices~\cite{Richter-Was:2018lld}
of reconstructed $2 \to 2$ effective Born frames of Collins-Soper~\cite{Collins:1977iv}
and of Mustraal~\cite{Berends:1983mi} type tend to lead to biggest differences.
We can  use these frames both for event-reweighting and for observable build.

\section{ Numerical results} 
\label{sec:numerical}

Numerical results presented below are based on events generated with {\tt Pythia 8.3}~\cite{Bierlich:2022pfr} 
for the $pp$ collisions at 13 TeV, using $q \bar q \to Z/\gamma^* \to \tau \tau $ matrix element, with invariant-mass range of $\tau\tau$ pair $m_{\tau\tau}$= 65-150 GeV.
The $\tau$ decays $\tau^\pm \rightarrow \rho^\pm \nu_\tau$ were modeled with the {\tt Tauola} decay 
library~\cite{Jadach:1993hs} with no spin correlations in the decaying $\tau$
pair. In total, we have  about $ 10^6$ events.
Then, the correlations were added using weight calculated with {\tt TauSpinner} program discussed
in Section~\ref{sec:algorithm}, both for effects from the correlations in the SM (EW corrections included) 
 and in the SM+NP models.
In Fig.~\ref{Fig:mtautau_costheta} distributions of the invariant mass $m_{\tau\tau}$ and $\cos\theta$ 
for the generated sample are shown. 
Using the weight calculated with {\tt TauSpinner}, the distributions are shown in the SM with the EW effects included.

\begin{figure}
  \begin{center}                               
{
   \includegraphics[width=7.2cm,angle=0]{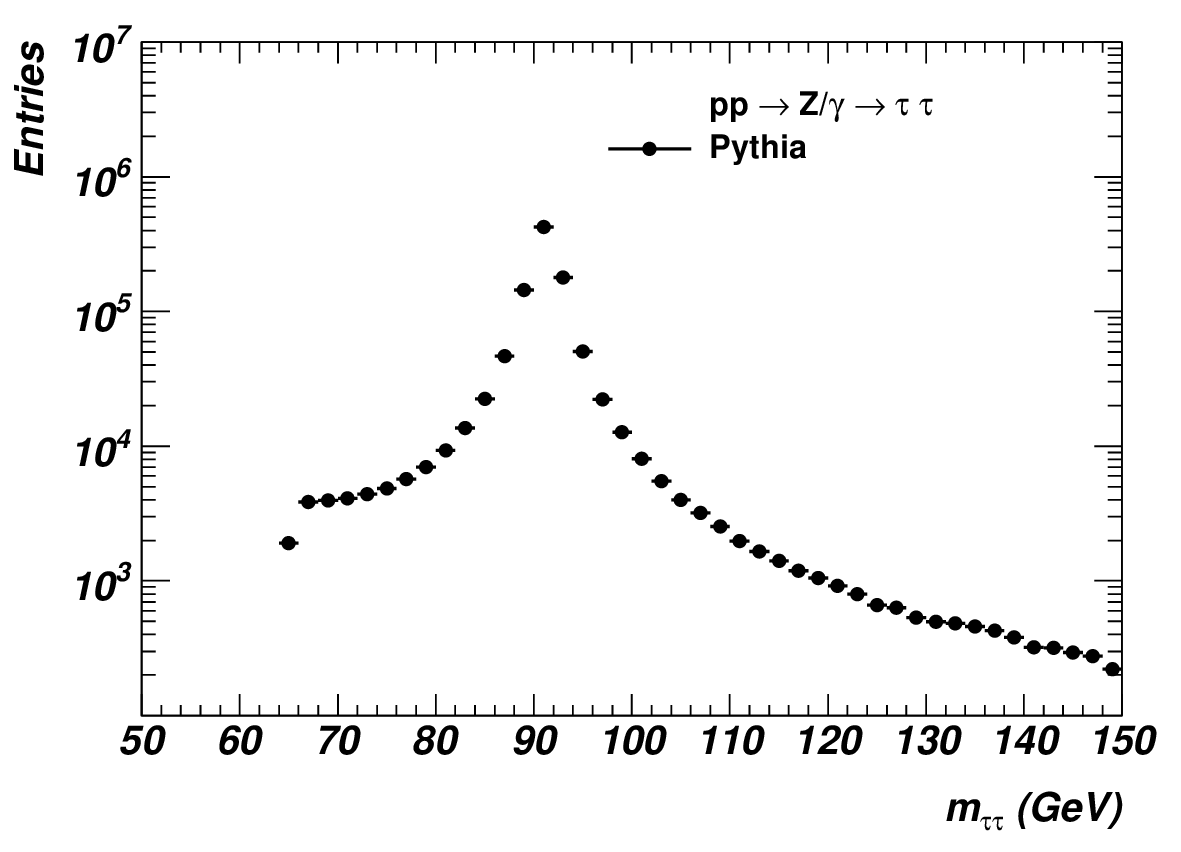}
   \includegraphics[width=7.2cm,angle=0]{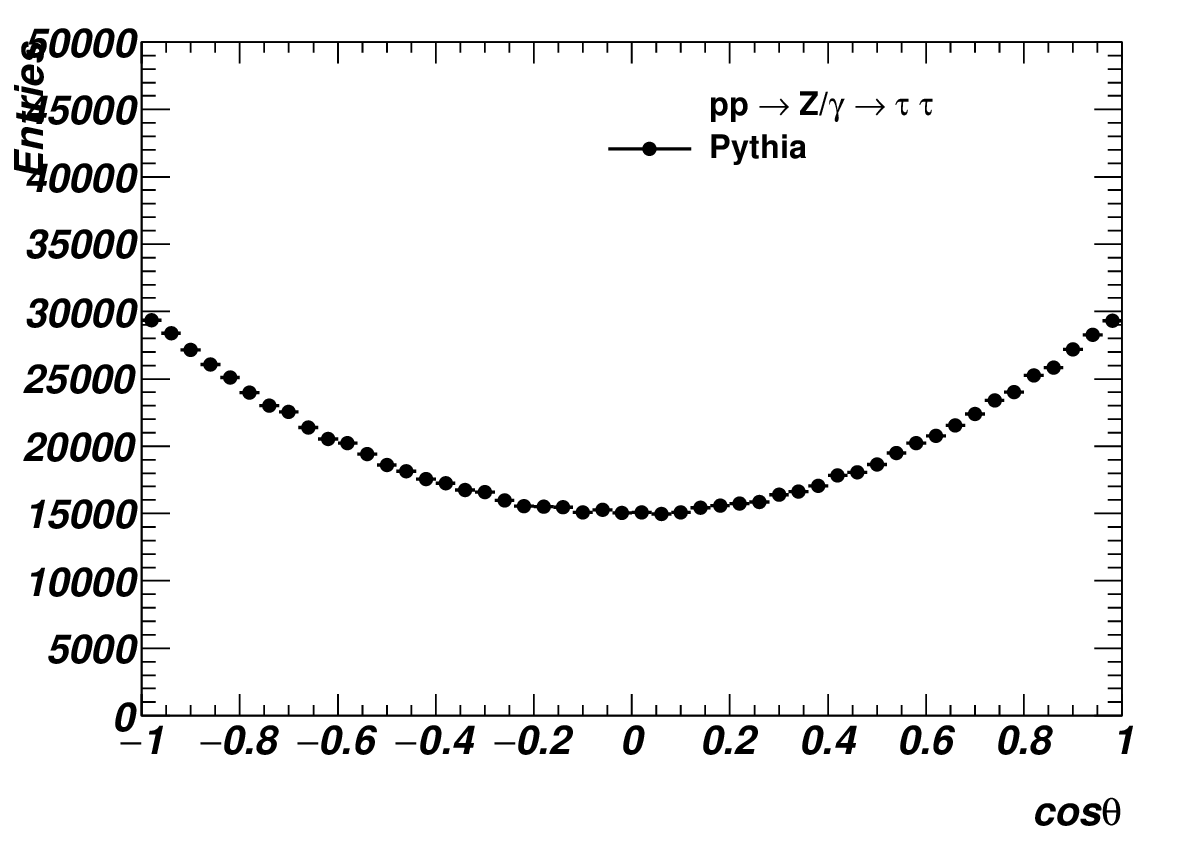}
}
\end{center}
\caption{Distribution of $m_{\tau\tau}$ and $\cos\theta$ of generated events 
from $q \bar q \to \tau\tau$ process in $pp$ collisions at 13 TeV centre-of-mass energy.
\label{Fig:mtautau_costheta} }
\end{figure}

In the following, we discuss numerical results for elements of spin-correlation matrix  $r_{ij}$
and identify which component may be most sensitive and provide some evidence of NP.
Then, we move to discussing impact on a few kinematical variables, typically studied in the
experimental analyses. The aim is to quantify effect from spin correlations as present in the SM and then
from NP extensions for $\Phi = \pm 0.1$, $A=0.1$ or $B=0.1$ and $X=0.1$ or $Y=0.1$.
Impact on the integrated cross-section due to the introduced NP effects is at the level of a few per mille at 
the $Z$-boson pole, and we do not discuss it here, as we are mostly interested in the spin-correlation effects.

\subsection{\texorpdfstring{Spin-correlation matrix elements $r_{ij}$}{}}
\label{sec:Rij}

To simplify discussion we use as a reference the SM with $\Phi = 0.0$, $A=B=0$ and $X=Y=0.0$. 
In particular, it is assumed that for the application, dipole moments due to QED loop corrections
are small and will be dropped out when discussing NP effects.
Therefore for numerical results we compare the SM predictions with four different settings for NP models: 
\begin{equation}
\label{eq:NP_settings}
(i) \;  \Phi=+0.1, \quad \; (ii) \; \Phi=-0.1, \quad \; (iii) \;  X=0.1, \, Y=0, \quad \;  (iv) \; X=0, \,  Y=0.1. 
\end{equation} 
We studied also non-zero values $A=0.1$ or $B=0.1$ in the $\gamma \tau \tau$ vertex, but the effect 
was very small and the corresponding plots are not included here. 

The distributions of $r_{xx}$ and $r_{xy}$, relevant for the transverse spin correlations, are shown 
in Fig.~\ref{Fig:rxx_rxy_SM} as functions of $m_{\tau\tau}$. 
The element  $r_{xx}$ is close to 0.5 around $Z$ peak, decreasing
fast below and above. There is almost no difference between IBA and BA. 
The non-zero element $r_{xy}$ in the BA is due to $Z \gamma$ interference term only, 
and is at the level of -0.015  (close to reported value in the case of $e^+e^-$collisions at 
LEP ~\cite{ALEPH:1997wux}), while in the IBA, which includes the EW form-factors, it
is at the level of -0.002.

In Fig.~\ref{Fig:rxx_rxy_mtautau} we show the similar distributions, but now comparing the SM case
in the IBA with the one, where NP effects are introduced as well.
The element $r_{xx}$ is almost not sensitive to the phase-shift $\Phi = \pm 0.1$, while $r_{xy}$
shows expected sensitivity, reaching the value of about $\pm 0.1$ for $m_{\tau \tau} = 80$ GeV.
The effect of NP model, realized with $X =  0.1$, introduces sizable increase of $r_{yy}$.   
{Element $r_{xx}$ also depends on $X$, however very weekly and only due to  
$X$-dependence of $R_{tt}$, which enters    
the definition $r_{xx} = R_{xx}/R_{tt}$, see~\cite{Banerjee:2023qjc}. This 
effectively induces difference between TT and NN correlation elements $r_{yy}$ and $r_{xx}$.}
Finally, as is seen from Fig.~\ref{Fig:rxx_rxy_mtautau}, the NT correlation element $r_{xy}$ 
is sensitive to the weak electric form-factor ${\rm Re}(Y)$. 

Of course, we should note that the values $X =  0.1$,  $Y =  0.1$ 
are a few orders of magnitude larger than the experimental limits on the weak dipole moments
obtained in Ref.~\cite{ALEPH:2002kbp}.

Distribution of  $r_{tz}$, relevant for the longitudinal polarisation, is shown  
as a function of $m_{\tau\tau}$ in Fig.~\ref{Fig:rtz_mtautau}. 
The element $r_{tz}$ is almost insensitive to the phase-shift $\Phi = \pm 0.1$, or $Y=0.1$,
while it is sensitive to the weak anomalous magnetic moment $X=0.1$, leading to a possible 
effect on the $\tau$-polarisation measurement
and related measurement of $\sin \theta_W$ in the $Z \to \tau \tau$ process.

\begin{figure}
  \begin{center}                               
{
   \includegraphics[width=7.2cm,angle=0]{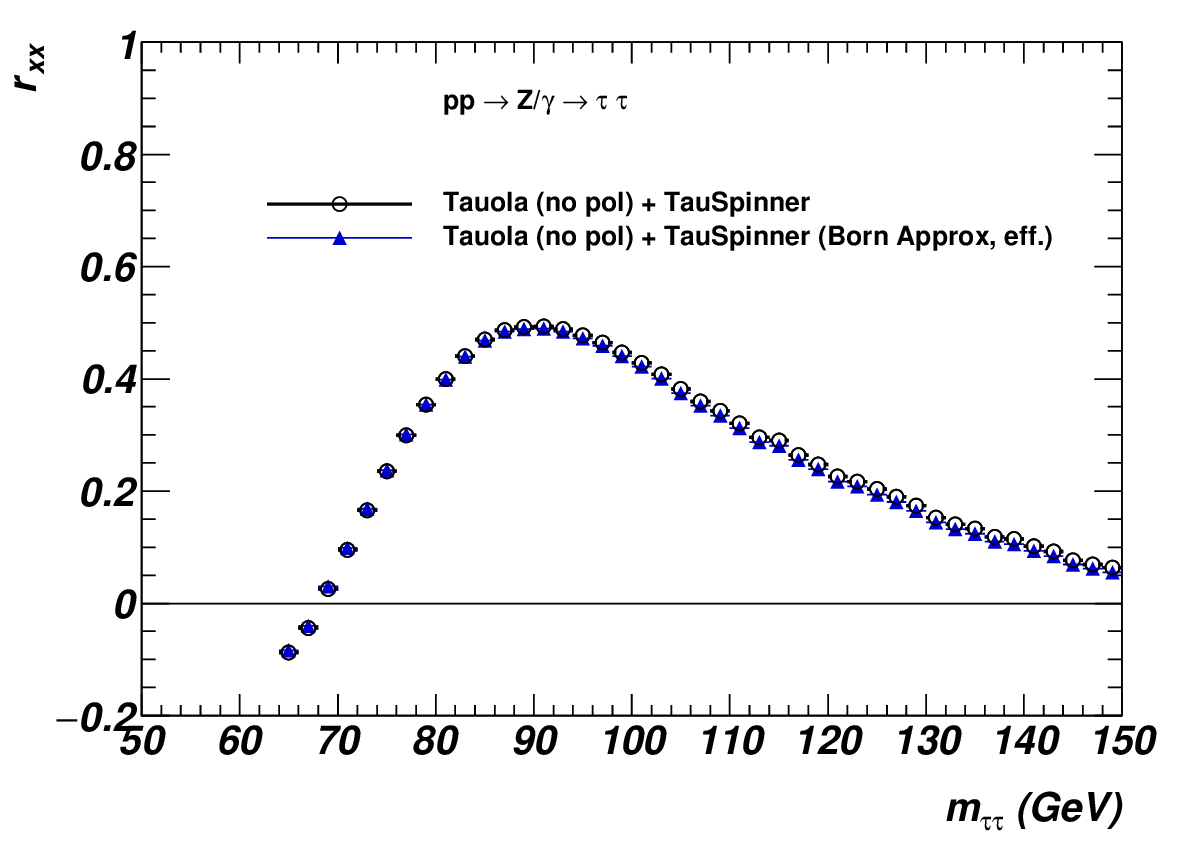}
   \includegraphics[width=7.2cm,angle=0]{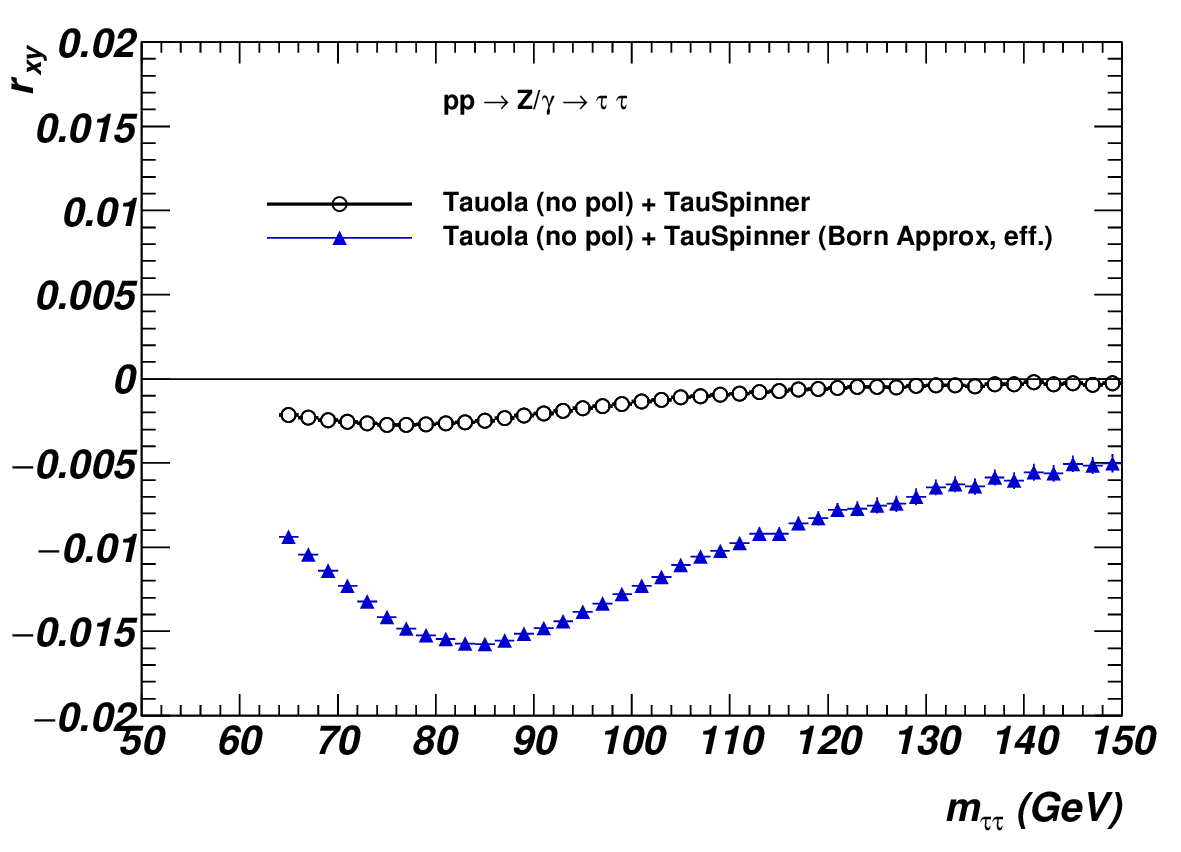}
}
\end{center}
  \caption{Distribution of $r_{xx}$ (left plot) and  $r_{xy}$ (right plot) as a function of invariant 
mass $m_{\tau\tau}$. Compared are the SM predictions using ${\cal M}^{IBA}$ (black open circles) 
and ${\cal M}^{BA}$  with effective couplings (blue triangles).
 \label{Fig:rxx_rxy_SM} }

   \begin{center}                               
{
   \includegraphics[width=7.5cm,angle=0]{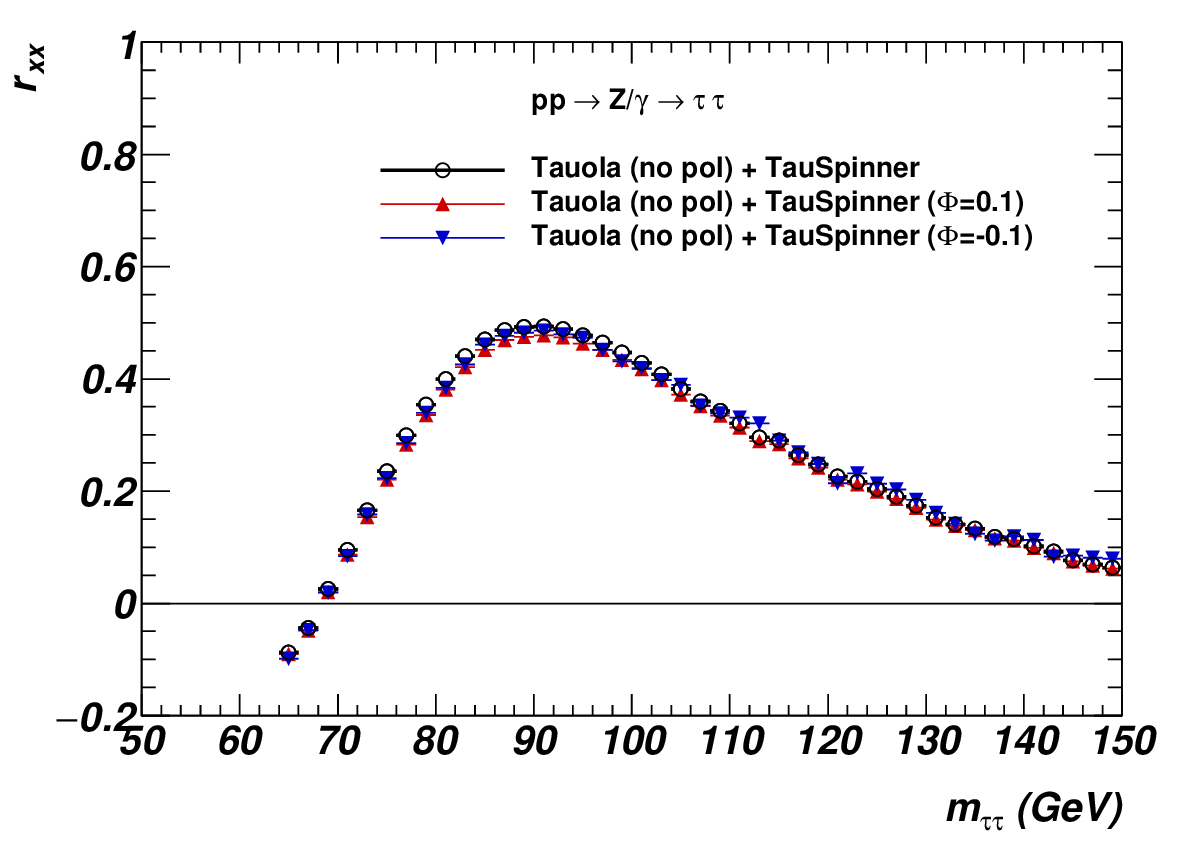}
   \includegraphics[width=7.5cm,angle=0]{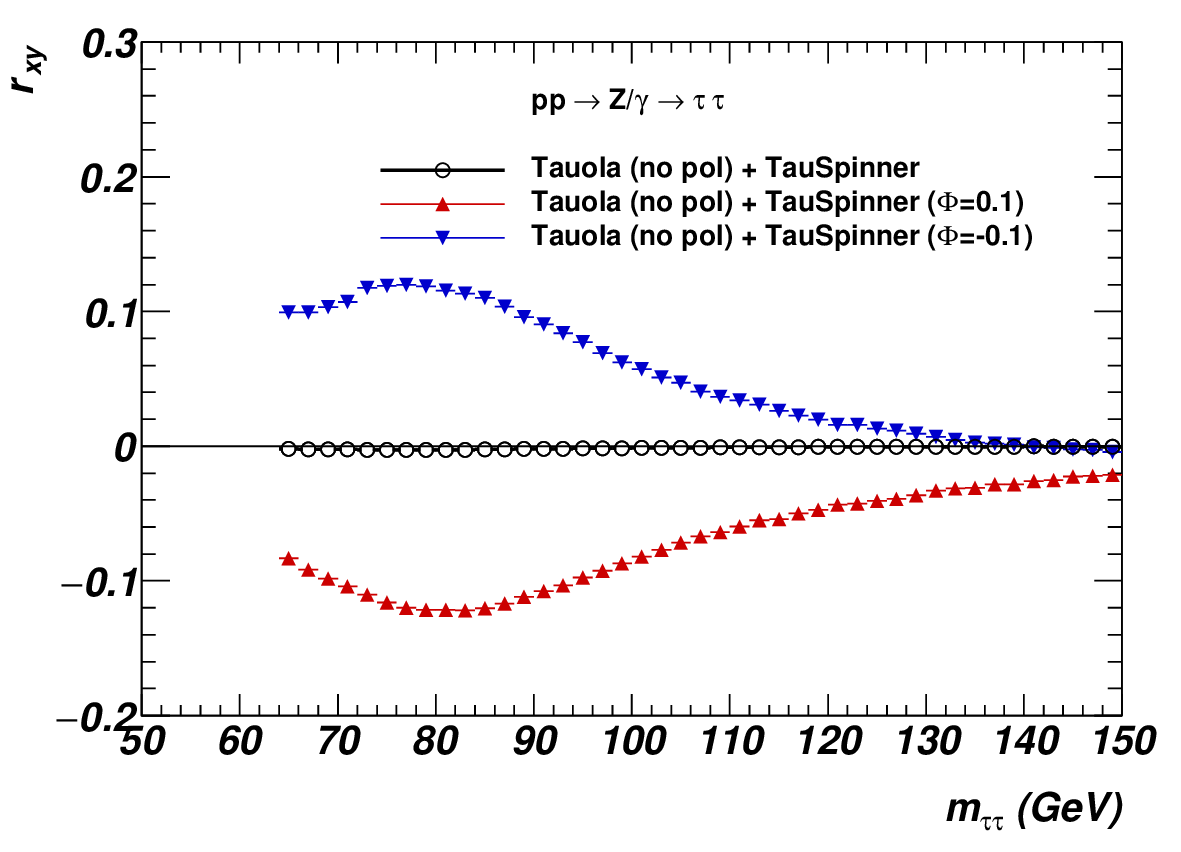}
   \includegraphics[width=7.5cm,angle=0]{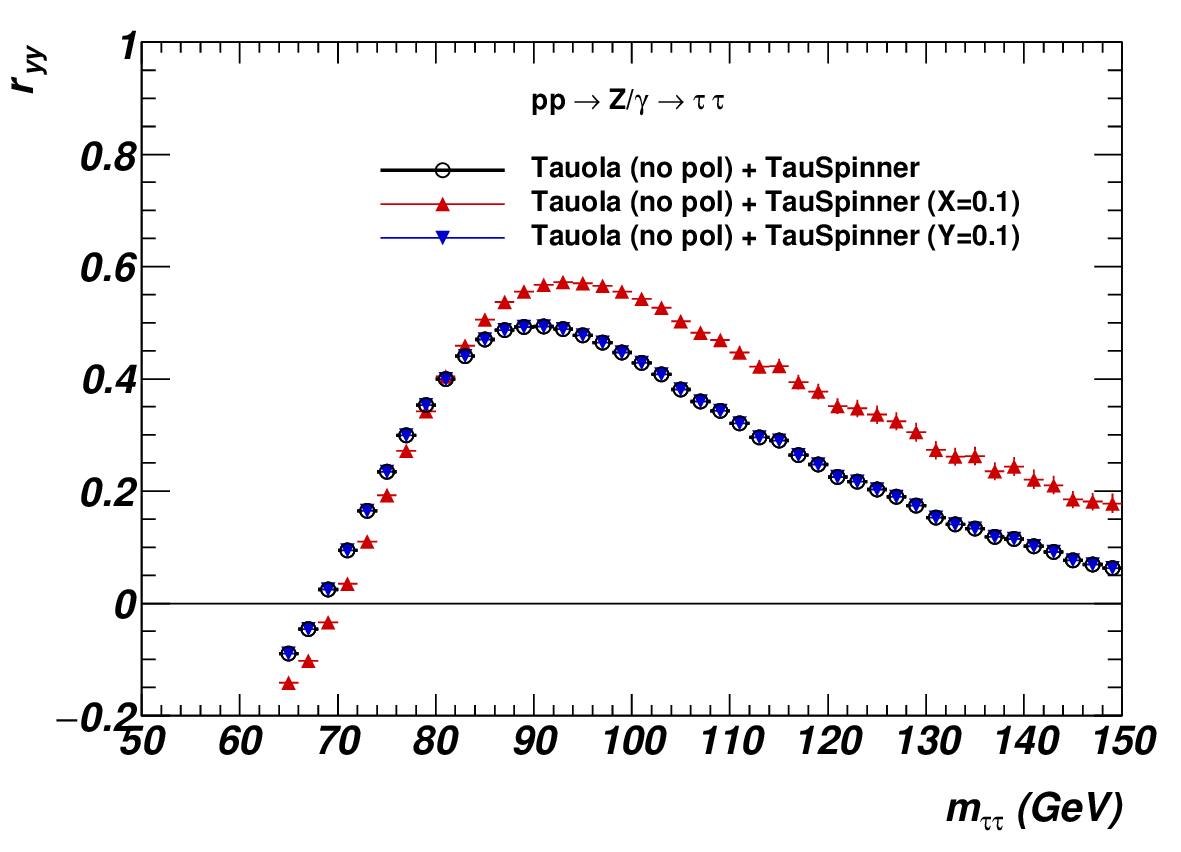}
   \includegraphics[width=7.5cm,angle=0]{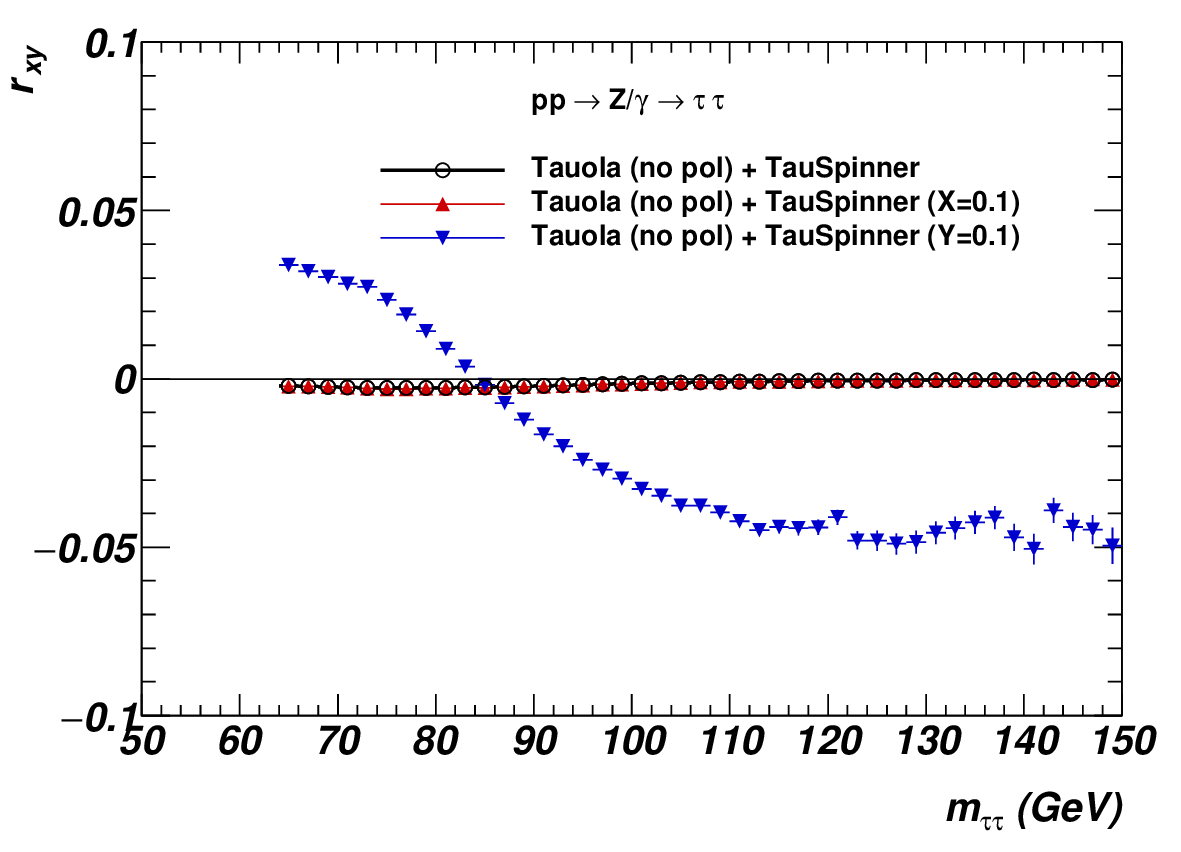}
}
\end{center}
\caption{Distribution of spin-correlation elements $r_{xx}$, $r_{xy}$ and $r_{yy}$  as a function 
of invariant mass $m_{\tau\tau}$.
Compared are the SM predictions using ${\cal M}^{IBA}$ (black open circles) and 
NP ones (red and blue triangles) with $\Phi= \pm 0.1$, $X=0.1$ or $Y=0.1$. 
Left column: $r_{xx}$ (top) and $r_{yy}$ (bottom); right column: $r_{xy}$.
 \label{Fig:rxx_rxy_mtautau} }
\end{figure}

\begin{figure}
  \begin{center}                               
{
   \includegraphics[width=7.2cm,angle=0]{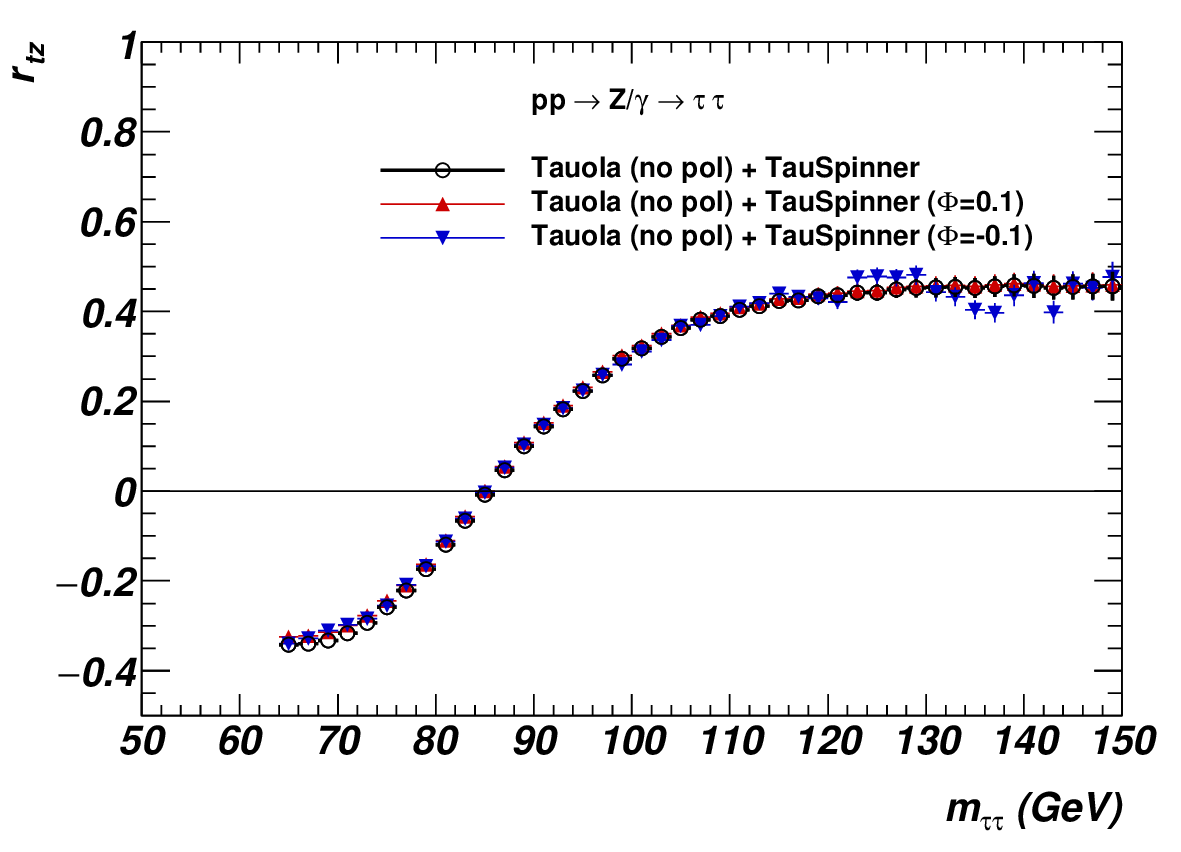}
   \includegraphics[width=7.2cm,angle=0]{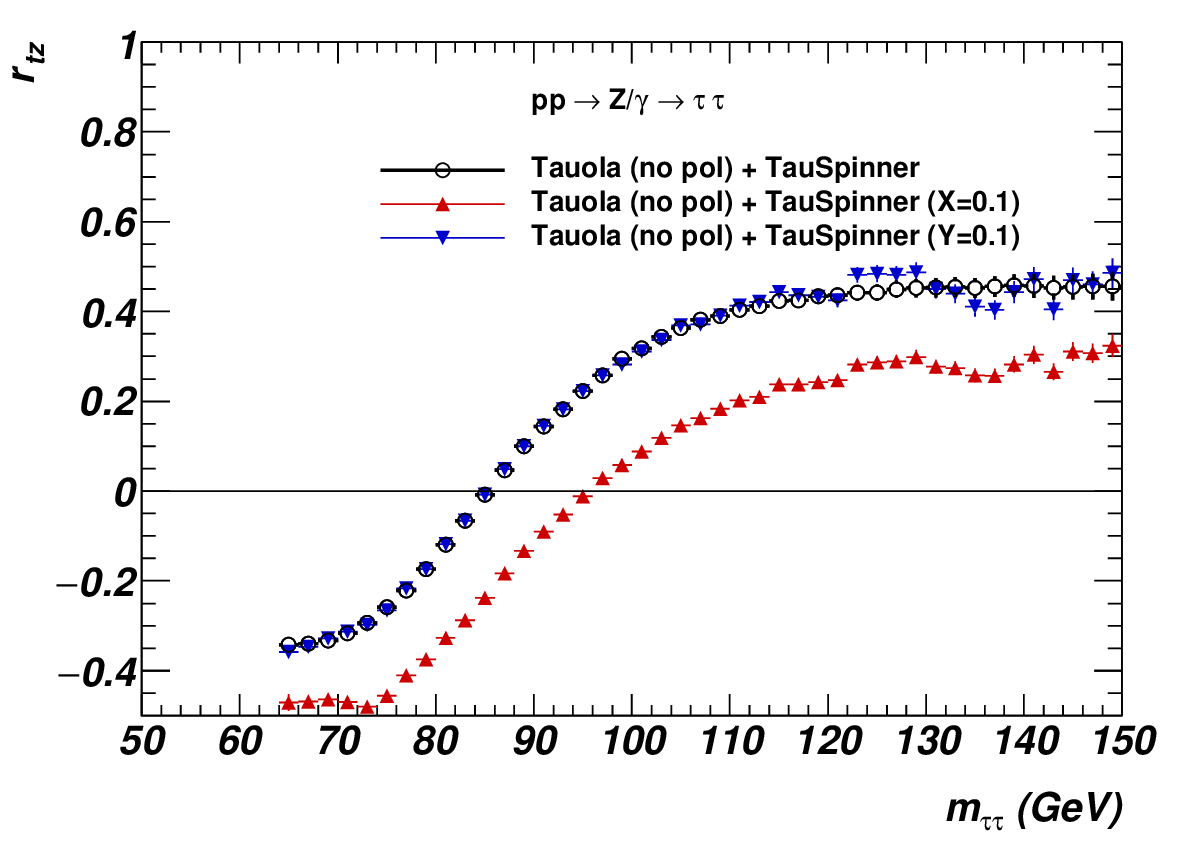}
}
\end{center}
  \caption{Distribution of $r_{tz}$ as a function of invariant mass $m_{\tau\tau}$.
    Compared are the SM predictions using ${\cal M}^{IBA}$ (black open circles) and NP ones (red and blue triangles):
    $\Phi = \pm 0.1$  (left plot)  and  $X=0.1$ or $Y=0.1$  (right plot).
 \label{Fig:rtz_mtautau} }
\end{figure}

\subsection{\texorpdfstring{Spin effects in $\tau^\pm \to \rho^\pm \nu_\tau$  decay channels}{}}
\label{sec:rhorho}

Polarisation and spin correlation effects are illustrated for a few experimental sensitive kinematical
variables listed below:
\begin{itemize}
\item
ratios $E_{\rho}/E_{\tau}$ and $E_{\pi^\pm}/E_{\rho}$, which are sensitive to longitudinal polarisation;
\item
ratio of invariant mass of $\rho^+\rho^-$ system to that of $\tau^+\tau^-$ system, 
$m_{\rho\rho}/m_{\tau\tau}$, which is sensitive to longitudinal spin correlations;
\item
  two acoplanarity angles calculated from kinematics of  decay products of the $\tau$ decays:
  $\Psi$ and $\phi^*$, which are sensitive to TT and TN spin correlations.
\end{itemize}

The definition of $\Psi$ distribution is based on the LEP publications~\cite{ALEPH:1997wux, DELPHI:1997ssw}.
The direction of the initial quark  (beam) is assumed to be along z-axis in the laboratory frame.
The four-vectors of $\tau$-decay products and beam vector are boosted to the rest frame of the $\rho^+ \rho^-$ 
system. In this frame, four-vectors are rotated around z- and y-axes such that $\pi^-$ is along z-axis, and
$\pi^+$ lies in the xz-plane. The $\Psi$ angle is defined as an angle between the beam direction and 
$\pi^-$ direction.

The definition of the angle $\phi^*$ was proposed for the Higgs CP measurement~\cite{Desch:2003rw} and
was used for the Higgs-boson CP measurement at the LHC~\cite{ATLAS:2022akr, CMS:2021sdq}. 
It is calculated as an angle
between the planes spanned on $\pi^{\pm}, \pi^0$ directions for each $\tau^\pm$ decay, calculated 
in the rest frame of the  $\rho^+ \rho^-$ system.
In case of $\phi^*$ angle, to preserve sensitivity to the transverse spin correlations,
sample has to be split depending on the angle $\alpha_\rho$, which is the angle between the beam axis and
plane spanned over $\pi^{\pm}, \pi^0$ from one decaying $\tau$ lepton, calculated in the laboratory frame
(splitting into regions: $\alpha_\rho <\pi/4$ and  $\alpha_\rho >\pi/4$). 

Both  $\Psi$ and  $\phi^*$ distributions show periodical cosine-shape modulation, once the sample is split in the 
two sub-samples depending on the sign of the product $y_{\tau^+} \cdot y_{\tau^-}$,  
where $y_{\tau^\pm}= (E_{\pi^\pm}-E_{\pi^0})/(E_{\pi^\pm}+E_{\pi^0})$.
If $y_{\tau^+} \cdot y_{\tau^-} < 0$, we shift $\Psi \rightarrow \Psi -\pi/2$ and $\phi^* \rightarrow \phi^* -\pi$.
This shift was integrated into definition of $\Psi$ and $\phi^*$ respectively, and then the final adjustment is made
to respect periodicity, i.e. with $\Psi$ being in the range $(-\pi, +\pi)$  and  $\phi^*$ in the range $(0, 2 \pi)$.
The periodical cosine-like characteristic in the $\Psi$ and $\phi^*$ distributions
is due to non-zero components $r_{xx}$, $r_{yy}$ of the spin-correlation matrix.
However, the phase of it is sensitive to non-diagonal $r_{yx}$ and $r_{xy}$ components.

Distribution of spin correlations sensitive kinematical observables in the SM  with 
${\cal M}^{IBA}$ (black open circles)
and NP model with $\Phi = \pm 0.1$ are shown in Fig.~\ref{Fig:Kinem_rhorho_TRfi}.
Those related to longitudinal polarisation 
and spin correlations,  $E_{\rho}/E_{\tau}$,  $E_{\pi^\pm}/E_{\rho}$ and  $m_{\rho\rho}/m_{\tau\tau}$, 
are  not affected.
The expected small shift in the phase of  $\Psi$ and $\phi^*$ distributions is observed, following impact on the
$r_{xy}$ distribution shown in Fig.~\ref{Fig:rxx_rxy_mtautau}.

The same set of kinematical distributions, but for NP models with $X=0.1$ or $Y=0.1$, is shown in 
Fig.~{\ref{Fig:Kinem_rhorho_XY}}. Distributions of
$E_{\rho}/E_{\tau}$,  $E_{\pi^\pm}/E_{\rho}$ and  $m_{\rho\rho}/m_{\tau\tau}$ show some effect
of changed $r_{tz}$ in case of $X=0.1$.  

One can also observe effect on  $\Psi$ and $\phi^*$ distributions, either canceling
cosine-like modulation in the $\Psi$ distribution, or changing the phase in the $\phi^*$ distribution. We expect some
effects due to changes in $r_{yy}$ in case of $X=0.1$, and in $r_{xy}$ in case of $Y=0.1$. 
But the interplay between changes in $r_{ij}$ and effects on distributions of sensitive acoplanarity angles 
seems quite non-trivial\footnote{We have presented numerical results with the use of
Collins-Soper frames for implementation of transverse spin and other effects, EW corrections 
and NP. We have also used Mustraal frame for that purpose. It was bringing no impact on the results, if
the $\tau$-pair transverse momentum was set to zero or was negligibly small. However some of our 
observables seemed to be less sensitive to implemented by {\tt TauSpinner} effect, if Mustraal frame 
was used.  As the Mustraal frame is more refined, we may expect that similar sensitivity will be if
actual data are used. Alternatively, an optimization of observables should be performed.
Unfortunately this requires major effort, possibly taking into account detector response and 
detection smearing, so we are bound to leave the topic for future work.}.

\begin{figure}
  \begin{center}                               
{
   \includegraphics[width=7.2cm,angle=0]{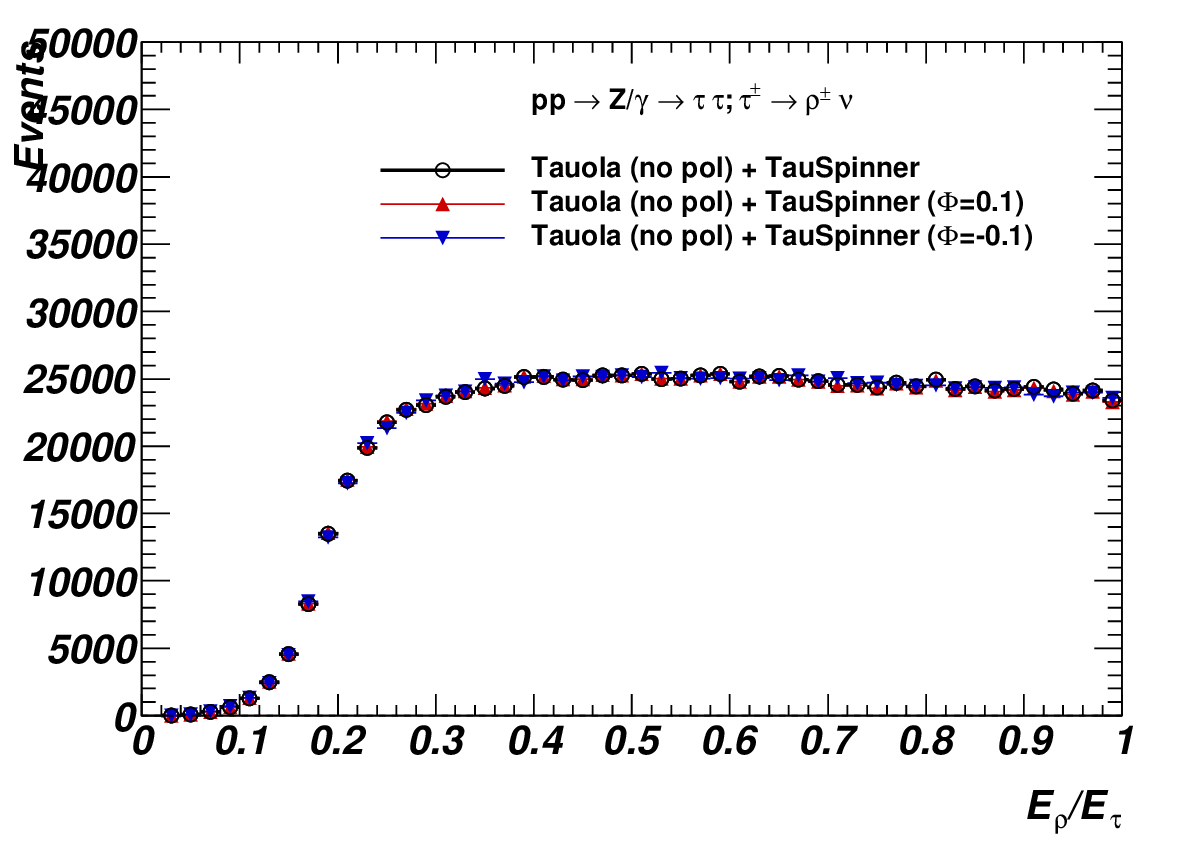}
   \includegraphics[width=7.2cm,angle=0]{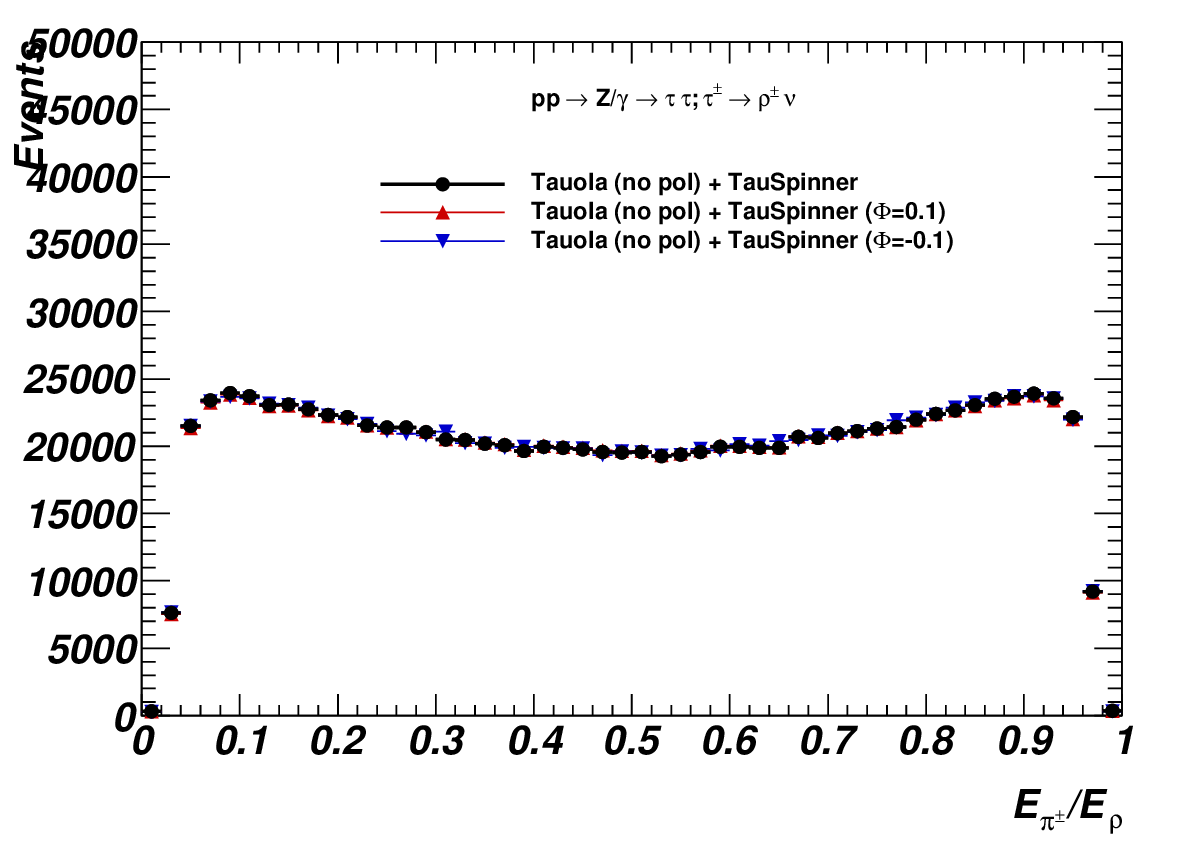}\\
   \includegraphics[width=7.2cm,angle=0]{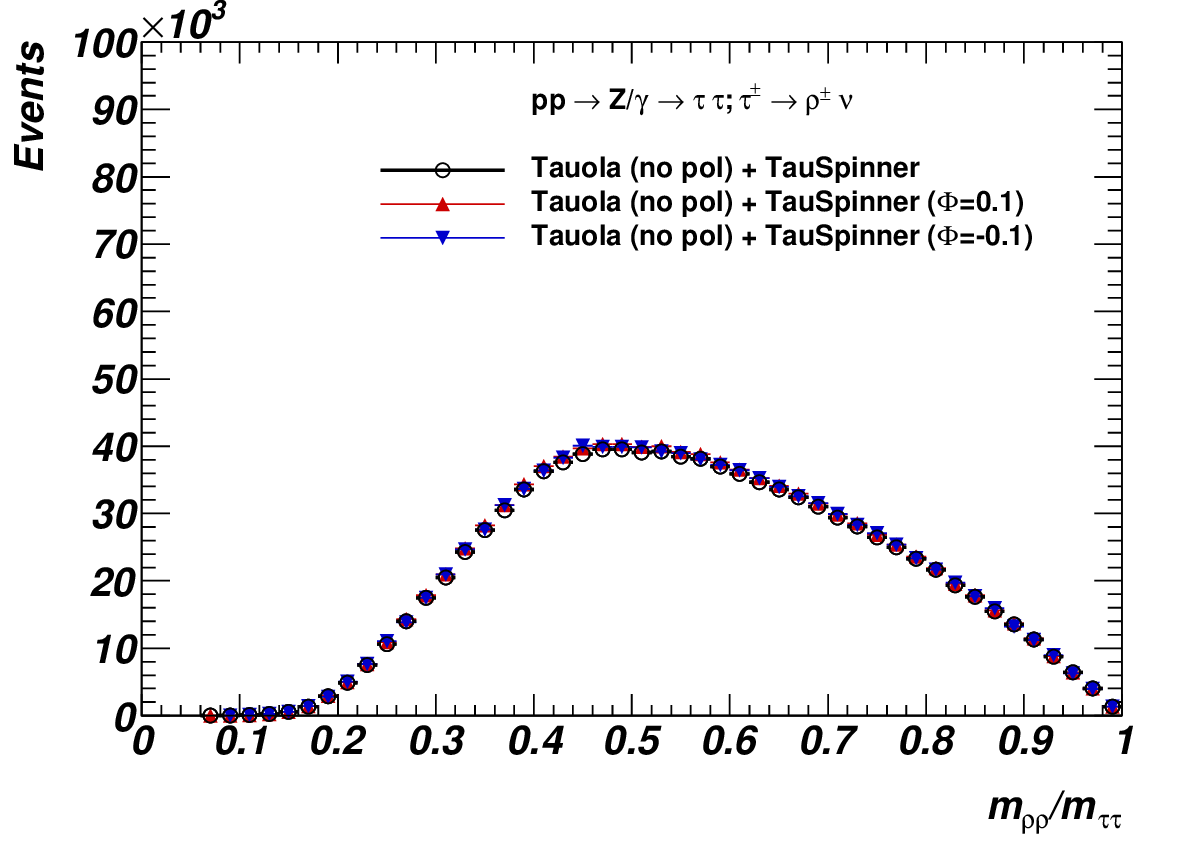}\\
   \includegraphics[width=7.2cm,angle=0]{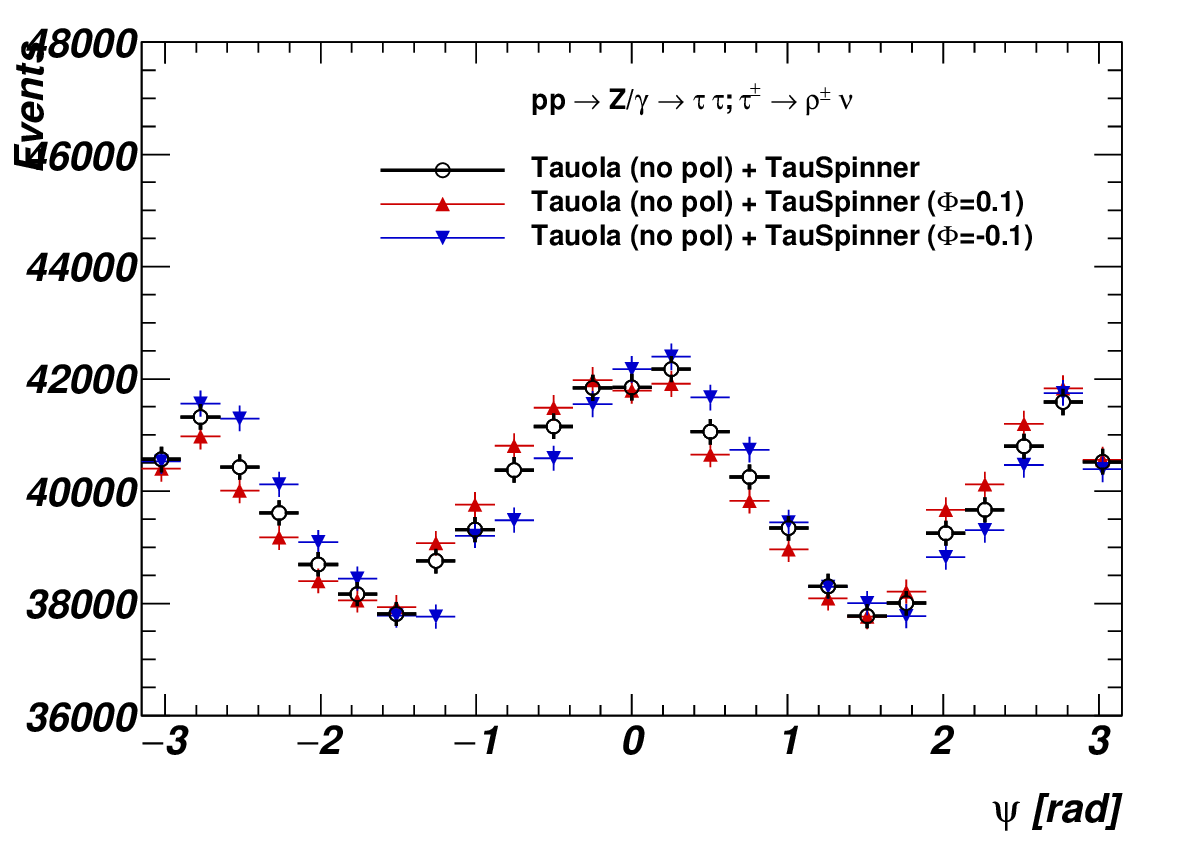}
   \includegraphics[width=7.2cm,angle=0]{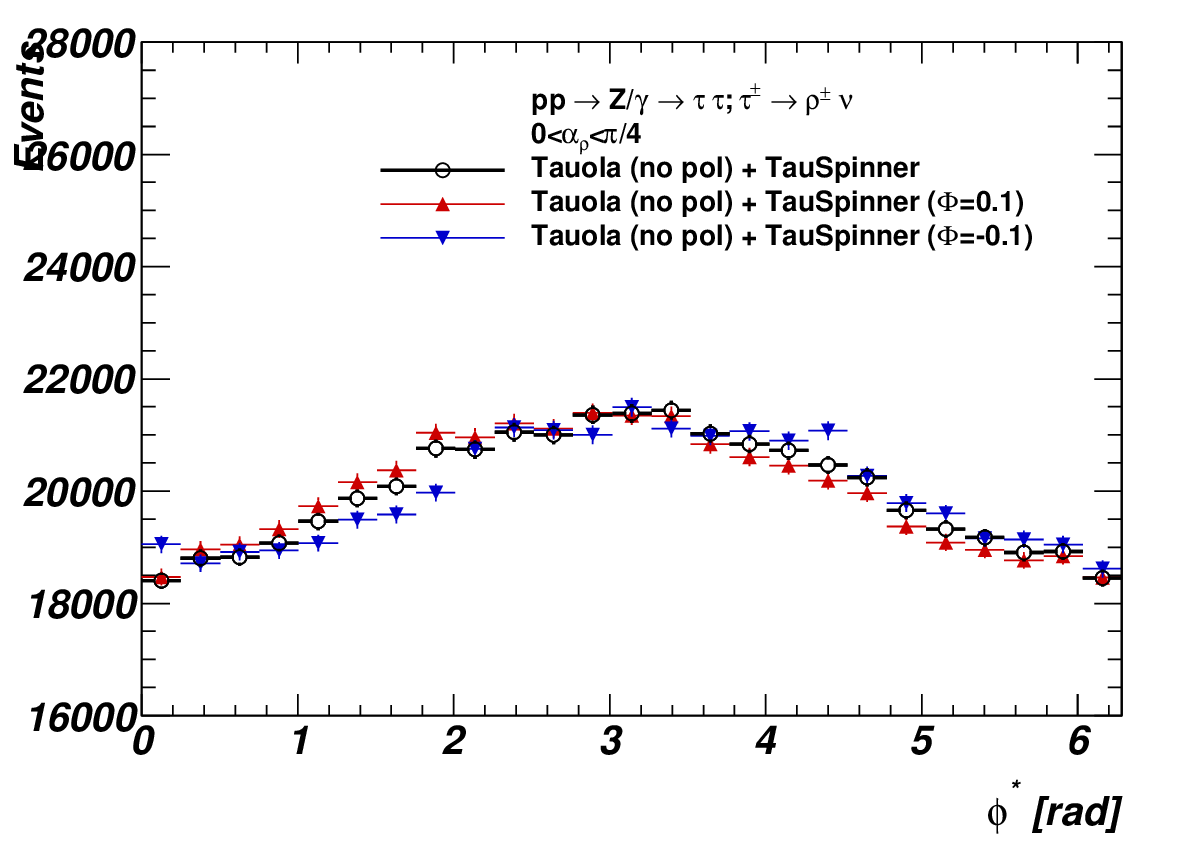}
}
\end{center}
\caption{Distribution of spin correlations sensitive kinematical observables.
Compared are the SM predictions ${\cal M}^{IBA}$ (black open circles) and NP ones (red and blue triangles) with 
$\Phi = \pm 0.1$. Both $\tau$ leptons decay via 
$\tau^\pm \to \rho^{\pm} \nu_\tau \to \pi^\pm \pi^0 \nu_\tau$.
 \label{Fig:Kinem_rhorho_TRfi} }
\end{figure}

\begin{figure}
  \begin{center}                               
{
   \includegraphics[width=7.2cm,angle=0]{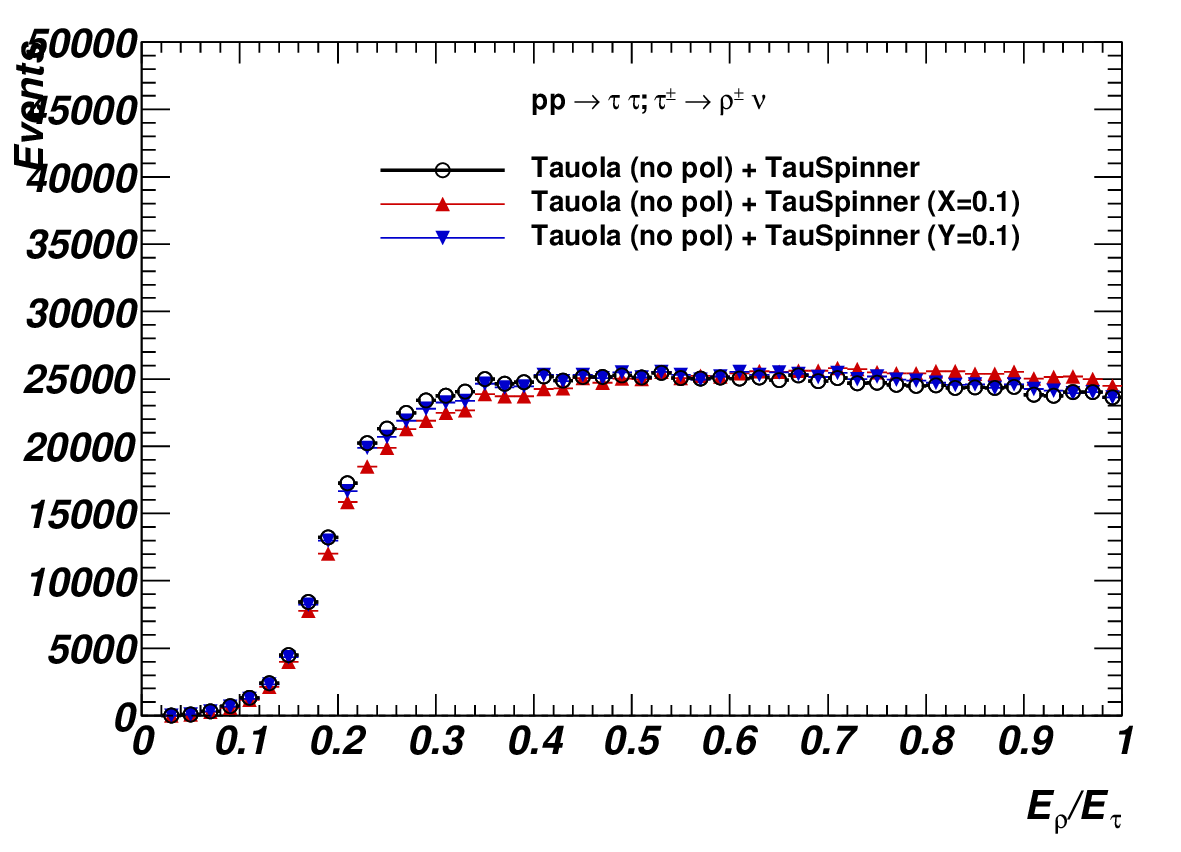}
   \includegraphics[width=7.2cm,angle=0]{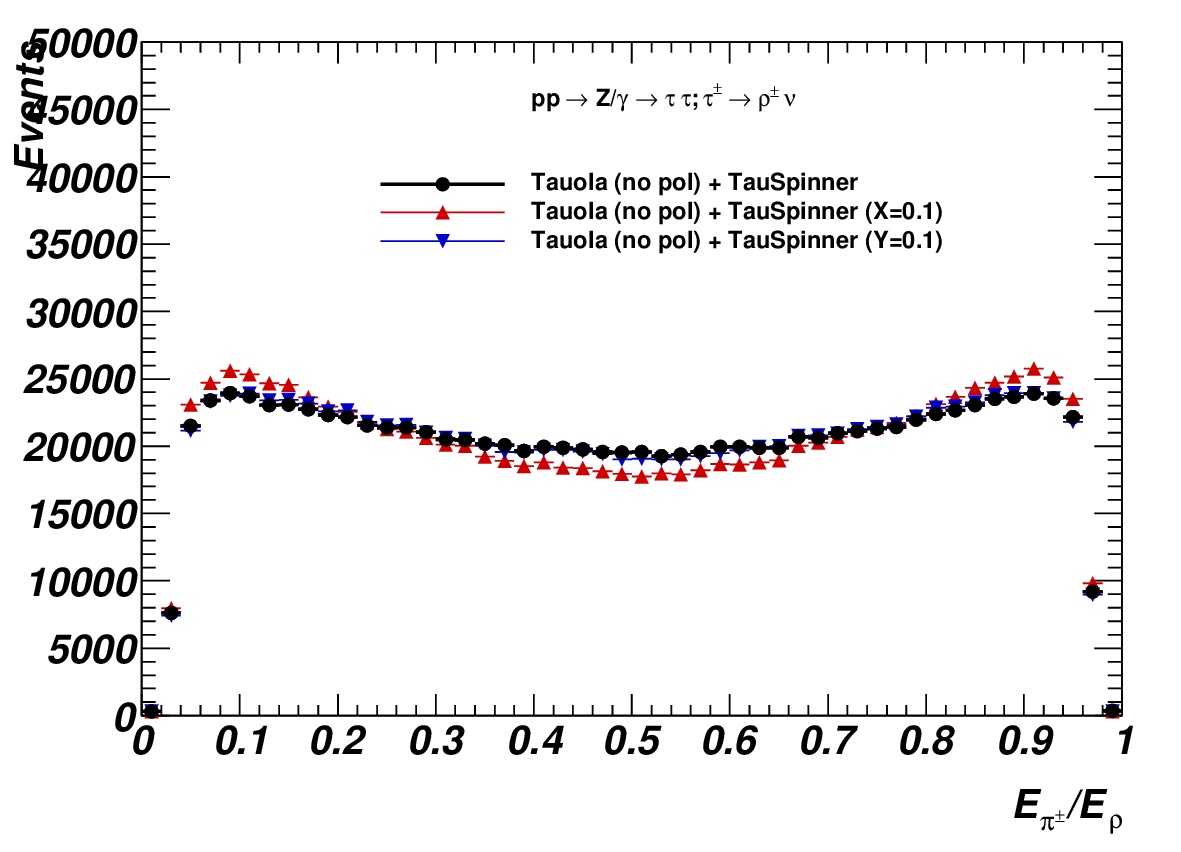}\\
   \includegraphics[width=7.2cm,angle=0]{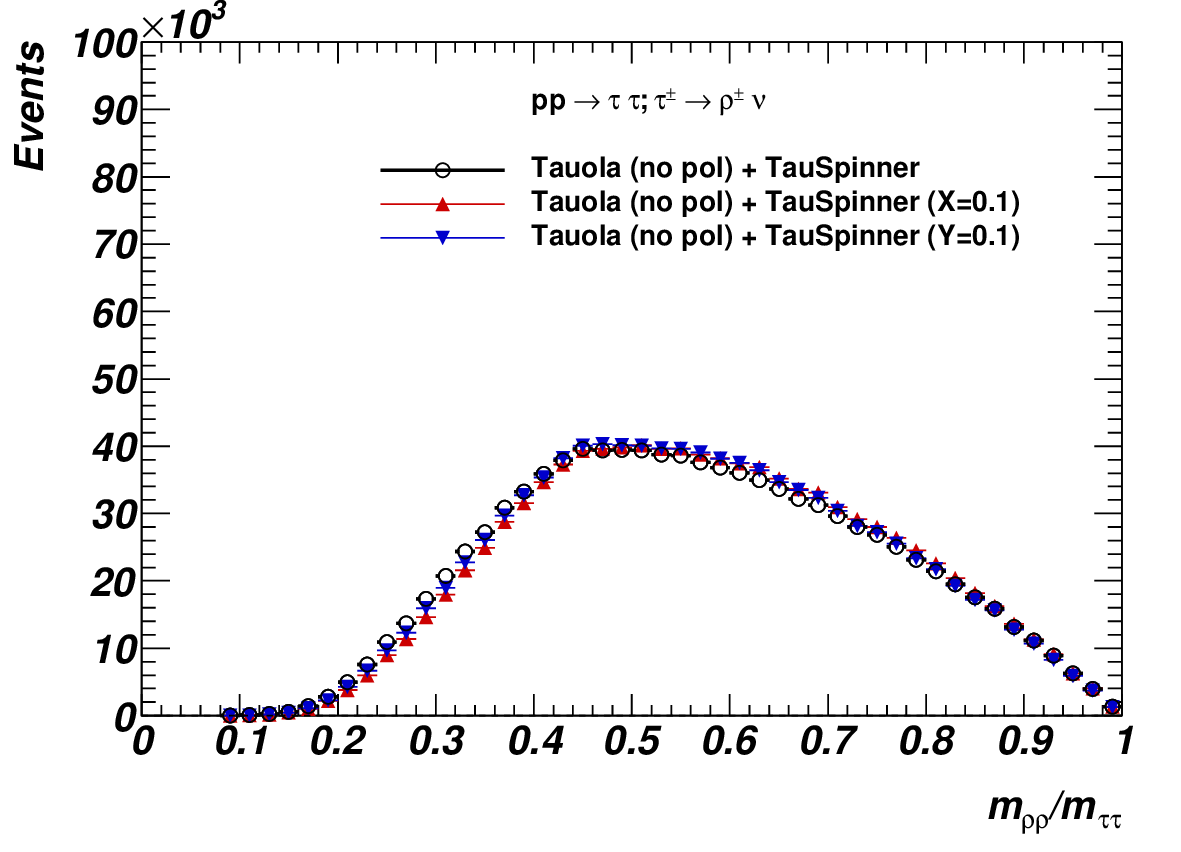}\\
   \includegraphics[width=7.2cm,angle=0]{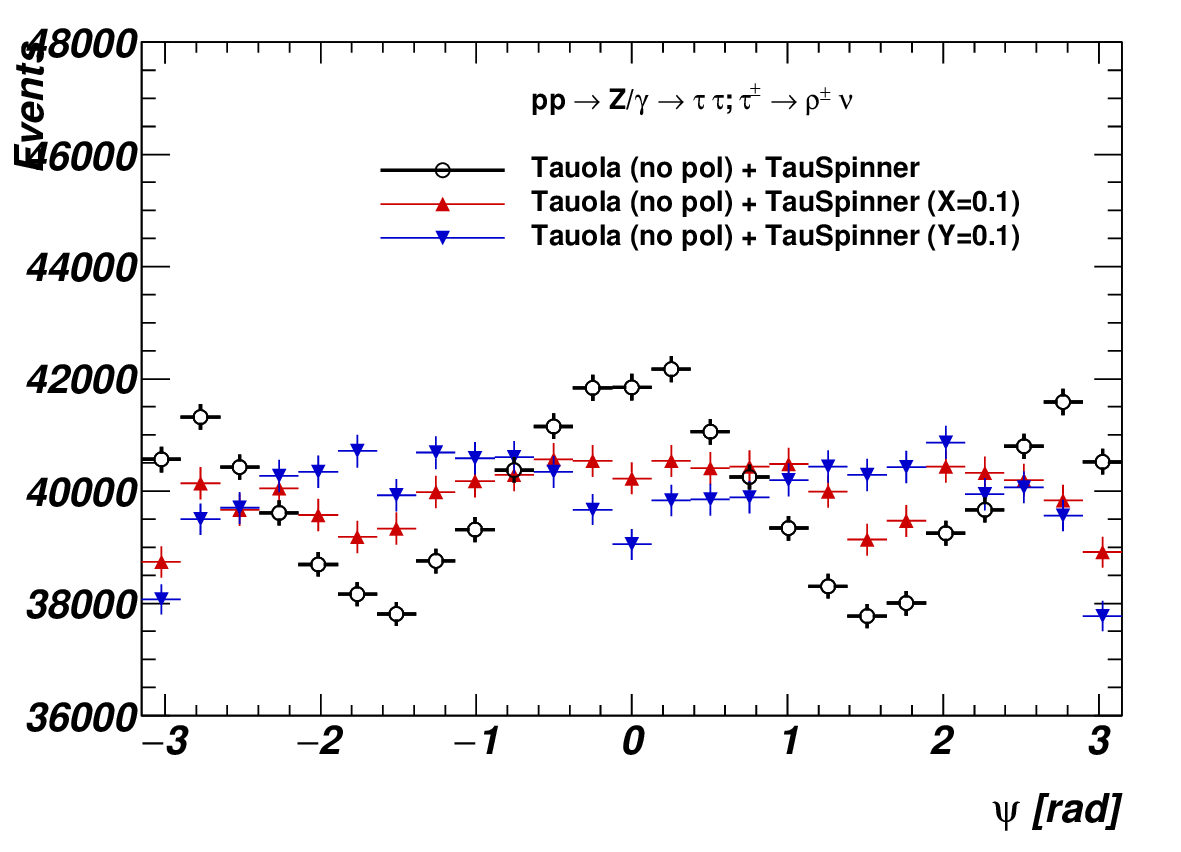}
   \includegraphics[width=7.2cm,angle=0]{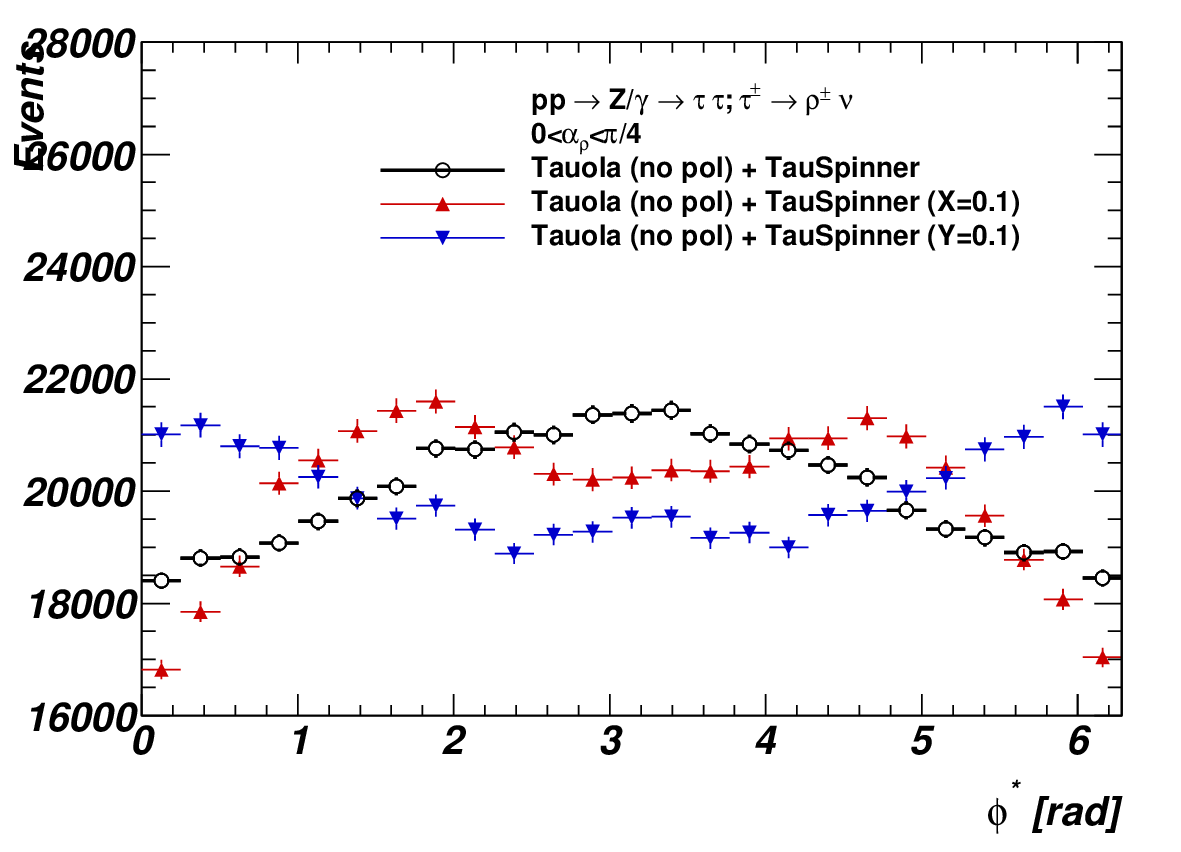}
}
\end{center}
\caption{Distribution of spin correlations sensitive kinematical observables.
Compared are the SM predictions ${\cal M}^{IBA}$ (black open circles) and NP ones (red and blue triangles) with 
$X=0.1$, or $Y=0.1$.  Both $\tau$ leptons decay via 
$\tau^\pm \to \rho^{\pm} \nu_\tau \to \pi^\pm \pi^0 \nu_\tau$.
 \label{Fig:Kinem_rhorho_XY} }
\end{figure}

\section{Summary and Outlook}
\label{sec:Outlook}

We have reviewed the dominant components of spin correlation matrices in $pp\to \tau \tau $ production with subsequent $\tau$ decays.
For that purpose, matrix element calculation was performed, including generic form-factors of potential New Physics.
Already the SM interactions in the Born Approximation feature vector and axial-vector couplings, distinct for incoming partons and outgoing $\tau$ leptons. Picture complicates further in the Improved Born Approximation 
because of the electroweak corrections which effectively introduce additional phases to the couplings, on top of the 
phase of the $Z$-boson propagator with respect to the propagator of the $\gamma$ exchange.
This requires attention as it may be expected to mimic effects of New Physics signature.
That is why, not only New Physics effects were installed into {\tt TauSpinner} event re-weighting algorithm, 
but it was assured that they can be studied simultaneously with the electroweak corrections.

After describing formalism, discussion on the numerical predictions for selected terms of spin-correlation matrix was presented. The focus was on the transverse spin correlations, which usually are not so often discussed in the literature,
in particular for $pp$ collisions.

Several sensitive distributions were studied, first for elements of spin correlation matrix, later for some semi-realistic distributions in which $\tau$ leptons decay into the channels 
$\tau^+ \to \rho^+ \bar{\nu}_\tau \to \pi^+ \pi^0 \bar{\nu}_\tau$,  \  
$\tau^- \to \rho^- \nu_\tau \to \pi^- \pi^0 \nu_\tau$. 
It was shown that exploring spin correlation can improve sensitivity of experiments to New Physics effects. On the other hand, if spin correlations are not taken into account, experimental selection may bias measurements 
of cross-sections and even mimic presence of New Physics. 


\vspace{0.4cm}

\centerline {\bf Acknowledgments}

\vspace{0.4cm}

This project was supported in part from funds of the National Science Centre, Poland,
grant no. 2023/50/A/ST2/00224 and of COPIN-IN2P3 collaboration with LAPP-Annecy.
A.Yu.K. is grateful to M.~Smoluchowski Institute of Physics of the Jagiellonian University 
and Institute of Nuclear Physics Polish Academy of Sciences for support. 

The majority of the numerical calculations were performed at the PLGrid Infrastructure
of the Academic Computer Centre CYFRONET AGH in Krakow, Poland.
\vspace{0.4cm}

\appendix

\section{\texorpdfstring{Transverse spin correlations at $Z$-boson pole }{}}
\label{app:ALEPH}

As a follow-up from the discussion in Section~\ref{subsec:SU2},
we can proceed with finding correspondence of $R_{ij}$, discussed here, and transverse-transverse (TT) and 
transverse-normal (TN)  spin correlations, discussed in Ref.~\cite{ALEPH:1997wux}.
Let us quote equation (1) for the cross-section used in~\cite{ALEPH:1997wux}
\begin{eqnarray}
  \frac{d \sigma(s_{\tau^-}, s_{\tau^+})}{d\Omega_{\tau^-}} = &&  \frac{1}{4 s} |P(s)|^2  \Bigl[\,  C_0 (1 + \cos^2\theta) + 2 C_1 \cos\theta  \nn \\
     && - D_0 (s^L_{\tau^-} -s^L_{\tau^+}) (1 + \cos^2\theta) - 2 D_1 (s^L_{\tau^-} -s^L_{\tau^+}) \cos\theta \nonumber \\
     && - C_0 (s^L_{\tau^-} s^L_{\tau^+}) (1 + \cos^2\theta) - 2 C_1 (s^L_{\tau^-}s^L_{\tau^+}) \cos\theta \nonumber \\
     && + C_2 ( s^N_{\tau^-} s^N_{\tau^+} -  s^T_{\tau^-} s^T_{\tau^+}) \sin^2\theta 
		+ D_2 ( s^N_{\tau^-} s^T_{\tau^+} +  s^T_{\tau^-} s^N_{\tau^+}) \sin^2\theta \Bigr], 
\label{eq:ALEPH_1997}				
\end{eqnarray}
where $P(s)$ is defined in \cite{ALEPH:1997wux}, and coefficients $C_i$, $D_i$ ($i=0,1,2$) are expressed through the couplings. The first line represents cross-section without $\tau$ spin,
the second line carries information about longitudinal polarisation of $\tau$, third line -- about longitudinal spin correlations and the last one -- about transverse spin correlations. The measured TT and 
TN spin correlations are defined by the ratios  $C_{TT} = C_2/C_0$ and $C_{TN} = D_2/C_0$, respectively.

In notations of our paper, these correlations are related to the elements $R_{12}$, $R_{21}$, 
$R_{11}$ and $R_{22}$ discussed in Subsection~\ref{sec:SM_Rij}.
For the clarity of the discussion we consider the IBA and keep only $Z$-boson contribution 
to $R_{ij}^{IBA}$ at the $Z$ pole ($s=M_Z^2$). For the TN correlations we find (label ``IBA'' is omitted below):
\beq
\label{eq:R12Z}
R_{12}^{(Z)} =  - 2 {\cal N} |Z_{\tau q}|^2 (a_q^2 + |v_q|^2)  a_\tau   |v_\tau^{\prime}| 
 \sin(\Phi_{v_\tau}) \sin^2 \theta,    \qquad \quad
{\cal N} \equiv \frac{e^4 M_Z^2}{64 s_W^4 c_W^4 \Gamma_Z^2 },  
\eeq
where $ \Phi_{v_\tau} $ stands for the phase of the $Z \tau \tau$ vector coupling, 
and the axial-vector coupling remains real and unchanged.     

Correspondingly, for the element $R_{44}^{(Z)}$ we have  
\beq
\label{eq:R44Z}
R_{44}^{(Z)} = {\cal N} |Z_{\tau q}|^2 \Bigl[ (a_q^2 + |v_q|^2) (a_\tau^2 + |v_\tau^{\prime}|^2) 
(1+ \cos^2 \theta) + 8  a_q  a_\tau  {\rm Re} (v_q)  |v_\tau^\prime| \cos(\Phi_{v_\tau}) \cos \theta \Bigr].
\eeq

Comparing these equations with Eq.~(\ref{eq:ALEPH_1997}) we obtain the coefficient $D_2$ of the TN spin correlations  
\beq
\label{eq:D2}
D_2 = - 2 {\cal N} |Z_{\tau q}|^2 (a_q^2 + |v_q|^2)  a_\tau  |v_\tau^{\prime} | \sin(\Phi_{v_\tau}) , 
\eeq 
and the coefficient $C_0$ 
\beq
\label{eq:C0}
C_0 = {\cal N} |Z_{\tau q}|^2 (a_q^2 + |v_q|^2) (a_\tau^2 + |v_\tau^{\prime}|^2)
\eeq
which allows one to calculate $C_{TN} = D_2/C_0$.  
Note, that our element $r_{12}^{(Z)} =  R_{12}^{(Z)}/R_{44}^{(Z)} $ 
does not coincide with $C_{TN}$ because of the angular factors included in the definition of $R_{12}^{(Z)}$ 
and $R_{44}^{(Z)}$ (see Eqs.~(\ref{eq:R12Z}) and  (\ref{eq:R44Z})).   

For the TT spin correlations we find
\beq
R_{11}^{(Z)} = {\cal N} |Z_{\tau q}|^2( a_q^2 + |v_q|^2)(|v_\tau^{\prime}|^2 - a_\tau^2) \sin^2 \theta,
\label{eq:R11Z}
\eeq
therefore the expression corresponding to coefficient $C_2$, defined in Eq.~(\ref{eq:ALEPH_1997}), reads as
\beq
\label{eq:C2}
C_2 =  {\cal N} |Z_{\tau q}|^2 ( a_q^2 + |v_q|^2)(a_\tau^2 - |v_\tau^{\prime}|^2) 
\eeq
which allows one to calculate $C_{TT} = C_2/C_0$.  
Similarly,  the element $ r_{11}^{(Z)} = R_{11}^{(Z)}/R_{44}^{(Z)} $ 
is not equal to $- C_{TT}$ because of the presence of angular factors in $R_{11}^{(Z)}$ and $R_{44}^{(Z)}$ 
(Eqs.~(\ref{eq:R11Z}) and  (\ref{eq:R44Z})).

The denominators in $C_{TN} = D_2/C_0$  and $C_{TT} = C_2/C_0$ cancel
out when calculating their ratio, and the expression for TN/TT correlations can be written as 
\begin{equation}
  \label{eq:CTT_CTN}
 \frac{C_{TN}}{C_{TT}} =\frac{D_2}{C_2} = - \frac{R_{12}^{(Z)}}{R_{11}^{(Z)}} =  
- \frac{ 2  a_\tau  |v_\tau^{\prime} |} 
 {(a_{\tau}^2 -  |v_{\tau}^\prime|^2)   }  \sin (\Phi_{v_\tau}) = 
- \frac{ 2  |a_\tau|  |v_\tau^{\prime} |} 
 {(a_{\tau}^2 -  |v_{\tau}^\prime|^2)   }  \sin (\Phi_{v_\tau} - \Phi_{a_\tau}), 
\end{equation}
where the last equality follows from the fact that $a_\tau$ is negative 
and its phase $\Phi_{a_\tau}$ is equal to $\pi$.

The explicit relation between $ \sin (\Phi_{v_\tau}) $ and phase-shift $\Phi$ can be calculated from 
Eqs.~(\ref{eq:v_tau_prime_expand}) and (\ref{eq:Phi_tau_sin_cos}).
With all EW form-factors set to 1.0, effectively going from IBA to BA as used in Ref.~\cite{ALEPH:1997wux}, 
the coupling $v_\tau^{\prime}$ receives phase only due to implicit phase-shift and 
\begin{equation}
  \label{eq:phi_tau_phi}
\sin (\Phi_{v_\tau}) = \frac{4 s_W^2 \sin (\Phi) }{ \big( 1 - 8 s_W^2 \cos (\Phi) + 16 s_W^4 \bigr)^{1/2}} , 
\qquad  \;
\cos (\Phi_{v_\tau}) = \frac{4 s_W^2 \cos(\Phi) -1 }{ \big( 1 - 8 s_W^2 \cos (\Phi) + 16 s_W^4 \bigr)^{1/2}}.
\end{equation}
With EW corrections restored, which introduce additional phase to the vector coupling,  Eqs.~(\ref{eq:phi_tau_phi}) become more involved and tedious, however they can be obtained directly from (\ref{eq:v_tau_prime_expand}) 
and (\ref{eq:Phi_tau_sin_cos}).

A more general definition of the TT and TN spin correlations in a wider range
of the invariant mass of the $\tau$ pair, including EW corrections, $\gamma$ exchange and $Z \gamma$ interference, can be obtained following formulas for $R_{ij}$ given
in Subsection~\ref{sec:SM_Rij}. It can also be expanded to include the $\tau$ dipole moments, using
expressions from~\cite{Banerjee:2023qjc}.
The present implementation in {\tt TauSpinner} allows one to include those extensions, 
see Appendix~\ref{app:TauSpinner}.

	
\section{TauSpinner: technical details}
\label{app:TauSpinner}

The most up-to-date documentation of the {\tt TauSpinner} algorithms can be found
in~\cite{Przedzinski:2018ett}, however over the years,  new development have been added,
as outlined in Section~\ref{sec:intro}.
With each of them usually came dedicated article describing improvement/extension of the physics aspects
covered, followed by example of numerical results and technical additions for the initialisation
and usage~\cite{Richter-Was:2018lld, Richter-Was:2020jlt, Banerjee:2023qjc, Korchin:2025vzx}.

To summarise those extensions which happened since~\cite{Przedzinski:2018ett}:
\begin{itemize}
\item
 New matrix element for $\bar q q \to Z/\gamma^* \to \tau^+ \tau^-$, with full control on longitudinal
 polarisation, longitudinal and transverse spin correlations in Born and Improved Born Approximation.
 It is discussed here and in ~\cite{Banerjee:2023qjc}.
\item
 Matrix element for $\gamma \gamma  \to \tau^+ \tau^-$ process with full control on longitudinal
 polarisation, longitudinal and transverse spin correlations in Born Approximation.
 This implementation is flexible enough to be used also for PbPb collision~\cite{Korchin:2025vzx}.
\item
 Extension to include in matrix elements listed above form-factors corresponding to dipole and
 weak dipole moments, in flexible enough form that can be used to simulate New Physics effects
 providing it has the same structure of couplings.
 For  $\gamma \gamma  \to \tau^+ \tau^-$ the dipole moments are included up to the fourth order,
 and for $\bar q q \to Z/\gamma^* \to \tau^+ \tau^-$ process in the lowest order, but in matrix element
 of Improved Born Approximation.
\item
 Extension to include  New Physics effects in form of phase-shift between axial-vector and vector couplings
 of $Z$ boson to $\tau$ leptons, as discussed in Subsection~\ref{subsec:SU2}.
\end{itemize}
   
While we will not make an attempt to include here complete {\it user guide}, we try to summarize
technical details relevant to what  discussed in this paper.

\begin{flushleft}
{ \bf General comments}
\end{flushleft}

Spin weight is calculated assuming that the $\tau\tau$ final state
is a product of either $\bar q q$ or $\gamma \gamma$ scattering.
The calculations of corresponding $R_{ij}$ elements of spin correlations matrix
are invoked and proportion with which each of the process contributed to the sum
in Eq.~(\ref{eq:parton-level}) depends on the parametrisation of the structure functions
$f(x_1, \ldots)$, $f(x_2, \ldots)$. Those represent probabilities of finding in the colliding beams
partons (quark, gluon or photon), carrying momentum fractions $x_1$, $x_2$.

In case of proton-proton beams, probabilities $f(x_1, \ldots)$, $f(x_2, \ldots)$ will be taken from
the PDFs library, e.g. parametrisations of~\cite{Klein:2016yzr, Xie:2021equ} which
include also the photon structure functions.
The parametrisation used should be indicated during initialisation.
For PbPb collision, library parametrising the photon flux is not easily available, user
will be required to set it up by hand, for example the $\gamma \gamma$ contribution
to $R_{ij}$ can be taken in fixed proportion to the summed over flavours $\bar q  q$ one.

\begin{flushleft}
{ \bf Package distribution}
\end{flushleft}

The tarball of the package can be downloaded from the web page \\
{\tt https://tauolapp.web.cern.ch/tauolapp/}

The {\tt TauSpinner} package is distributed in the same tarball as {\tt Tauola} package,
a library for simulating for $\tau$ decays, as they share several components of the code,
interfaces and tests.

The installation script is prepared and is located in main directory\\
{\tt tauola/install-everything.sh}\\
which is installing both  {\tt Tauola/TauSpinner} but also other packages needed for
execution of the code and/or examples:
{\tt HepMC}, {\tt LHAPDF}, {\tt PHOTOS}, {\tt MCTESTER}, {\tt Pythia}.
One can comment it out and provide link to already existing installation in ones own environment.

The examples of use with short README can be found in the directory\\
{\tt tauola/TauSpinner/examples} of the distributed tarball.
In particular,  the following files provide good starting point.
\begin{verbatim}
read_particles_from_TAUOLA.cxx
tau-reweight-test.cxx
\end{verbatim}

\begin{flushleft}
{ \bf Initialisation}
\end{flushleft}

The user is required to configure both  $\bar q q$ and  $\gamma \gamma$ scattering.
In case the  $\gamma \gamma$ scattering will overly contribute, details
of the configuration used for  $\bar q q$ process are not relevant.
In case the used PDF library does not provide parametrisation of photon structure function of the proton,
user is required to set also proportion of $\gamma \gamma$ process with respect to  $\bar q q$ one,
as used at generation {\tt GAMfraci} and for desired NP model  {\tt GAMfrac2i}.

To invoke most advanced matrix element implementation, for  both  $\bar q q$ and $\gamma \gamma$ processes
the flag  {\tt ifkorch = 1} is mandatory, as it invokes consistent flow of the $R_{ij}$ and final weight
calculations. Details of implementation of corresponding matrix elements was given in~\cite{Banerjee:2023qjc},
with extension to higher-order terms for electromagnetic dipole moments for $\gamma \gamma$ processes
presented in~\cite{Korchin:2025vzx} and allowing for phase-shift between axial-vector and vector couplings
as discussed in Section~\ref{subsec:SU2} in this paper.

The flag {\tt iqed = 0} switches off  SM component $A0$  of the magnetic dipole moment, then the
total $A$, $B$ values can be defined by the user.

When including  $\gamma \gamma$ process, at the initalisation step, user is required to provide values
of the dipole moments used for sample generation $ A0i, B0i, X0i, Y0i$ and those of the NP model for
which weight will be calculated $Ai, Bi, Xi, Yi$. Reinitialisation can be done on per-event basis, if
needed, and the $s$-dependence of $A(s), B(s), X(s), Y(s)$ can be handled during events processing.

Few options are available for configuration of EW corrections, which are invoked with {\tt keyGSW} flag.
They allow to switch on/off some of the form-factors, calculated with {\tt Dizet}
library~\cite{Bardin:1999yd}, more details can be found in~\cite{Richter-Was:2020jlt}.
The form-factors are tabulated in {\tt table.up} and  {\tt table.down} files.
The defaults are: {\tt keyGSW = 1}  (complete set of EW form-factors);
{\tt keyGSW = 0} (EW form-factors not used, ME with effective couplings).

The legacy implementation of matrix element for  $\bar q q \to Z/\gamma^* \to \tau^+ \tau^-$ process,
documented in~\cite{Przedzinski:2018ett}, is still available. It is invoked with {\tt ifkorch = 0} flag.
The default is Born Approximation with effective couplings and only longitudinal polarisation and
spin correlations included. The Improved Born Approximation, including also transverse spin correlations
is used automatically instead, if two files are available  from the run directory:
{\tt table1-1.txt}, {\tt table2-2.txt}. They  contain pretabulated $r_{ij}, \, i,j =x, y$
values calculated with {\tt SANC} program~\cite{Andonov:2008ga}, see more in~\cite{Przedzinski:2018ett}.
We have checked, that results for $r_{xx}, r_{xy}$ are consistent between legacy implementation using
{\tt SANC} tables and new implementation using {\tt ifkorch =1} and {\tt Dizet} tables are used now.

Choice of the frame for calculating components of spin weight is steered with flag {\tt FrameType}.
With choice {\tt FrameType=1}  Mustraal frame (NLO like) is used, with {\tt FrameType=0} (default)
the similar to  Collins-Soper frame is used. The frame type can be initialized invoking:
{\tt void setFrameType(int \_FrameType)}.

We expect, that in a standard usage,  initialisation of parameters will be performed once.
However, re-initialisation of the parameters is possible on event-by-event base.
One can change e.g. setting of EW corrections or dipole moments  and process the same event several times
for weight calculation.
The ratio of the weights calculated for two sets of parameters, can be  then used to estimate
interesting properties of NP models in the process of analysis.

Below snippet of the initialisation code is shown. The path to necessary  {\tt includes} can be found in 
{\tt tauola/TauSpinner/examples/tau-reweight-test.cxx}
\begin{verbatim}

// initialisation of main flow of TauSpinner 
   double CMSENE = 13000.0; // center of mass system energy
                            // used in PDF calculation. For pp collisions only
   bool Ipp   = true;       // for pp collisions, the only option implemented
                            // but gam gam events from PbPb events can be also
                            // processed, see details later 
   int Ipol   = 0; // are input samples polarized?
   int nonSM2 = 1; // are we using nonSM calculations? 
   int nonSMN = 0; // If we are using nonSM calculations we may want corrections
                   // to shapes only: y/n  (1/0)
   TauSpinner::initialize_spinner(Ipp, Ipol, nonSM2, nonSMN,  CMSENE);

// NEW: initialisation of the frame used for weights calculation
   setFrameType(1); // 1 for Mustraal, 0 for Collins-Soper

// initialisation for Dizet electroweak tables, here you specify location
// where the tables are present, default: in your run directory
// for more documentation see arXiv:2012.10997, arXiv:1808.08616  
   char* mumu="table.mu";
   char* downdown= "table.down";
   char* upup= "table.up";
   int initResult=initTables(mumu,downdown,upup);

// initialisation for EW parameters, Born Approximation with effective couplings
// for more documentation see  arXiv:2012.10997, arXiv:1808.08616 
   double SWeff=0.2315200;
   double DeltSQ=0.;
   double DeltV=0.;
   double Gmu=0.00001166389;
   double alfinv=128.86674175;
   int keyGSW=1;
   double AMZi=91.18870000;
   double GAM=2.49520000;
   ExtraEWparamsSet(AMZi, GAM, SWeff, alfinv,DeltSQ, DeltV, Gmu,keyGSW);
 
// initialisation for the qbar q ->tautau matrix elements
// dipole moments and weak dipole moments are set to 0.0 in the qbar q->tautau ME
// flags relevant here are:
   int ifGSW   = 1;  // use EW form-factors
   int ifkorch = 1;  // global flag to switch on ME calculations of arXiv:2307.03526
   int iqed = 0;     // iqed = 1  (if A0 at SM value to be added) 
   initialize_GSW(ifGSW, ifkorch, iqed, 0.0, 0.0, 0.0, 0.0, 0.0, 0.0, 0.0, 0.0);

  //  set multipliers of  (Rxx, Ryy, Rxy, Ryx)      
  setZgamMultipliersTR(1., 1., 1., 1. );

  // LEGACY: for using implementation of (Rxx, Ryy, Rxy, Ryx) as in arXiv:1802.05459
  // transverse spin correlations Rxx, Ryy, Rxy, Ryx read in  from tables 
  // table1-1.txt, table2-2.txt which should be placed in the run directory. 
  initialize_GSW(0, 0, 0.0, 0.0, 0.0, 0.0, 0.0, 0.0, 0.0, 0.0);


  // NEW: for using implementation of (Rxx, Ryy, Rxy, Ryx) as in arXiv:2307.03526 
  initialize_GSW(1, 1, 0.0, 0.0, 0.0, 0.0, 0.0, 0.0, 0.0, 0.0); // SM with Improved  Born 
  initialize_GSW(0, 1, 0.0, 0.0, 0.0, 0.0, 0.0, 0.0, 0.0, 0.0); // SM with Effective Born 

  // NEW: for using implementation of (Rxx, Ryy, Rxy, Ryx) as in arXiv:2307.03526 
  // adding magnetic and electric dipole moments and weak dipole moments
  // example initialisation, requires flags: ifkorch = 1; nonSM2 = 1; 
  double ReA = 0.1;
  double ImA = 0.0;
  double ReB = 0.0;
  double ImB = 0.0;
  double ReX = 0.0;
  double ImX = 0.0;
  double ReY = 0.0;
  double ImY = 0.0;
  initialize_GSW(1, 1, ReA, ImA, ReB, ImB, ReX, ImX, ReY, ImY); // with Improved  Born 

  // NEW: form-factors from  Dizet electroweak tables  can be scaled
  // here use form-factors as in  Dizet electroweak tables, scaled by 1.0
  // select option for EW form-factors
  int keyGSW = 1;  //default
  initialize_GSW_norm(keyGSW, 1.0, 0.0, 1.0, 0.0, 1.0, 0.0, 1.0, 0.0, 1.0, 0.0, 1.0, 0.0);

  // NEW: introduce shift between phase of axial and vector couplings of $Z$ to tau leptons
  // use scaling for this purpose, as described in Section 2.3
  double fi=0.1;
  double c=cos(fi);
  double s=sin(fi);
  initialize_GSW_norm(keyGSW, c, -s, c, s, 1.0, 0.0, c, s, 1.0, 0.0, 1.0, 0.0);

\end{verbatim}

The default setting used, if not overwritten during initialisation, can be found at the very top of the file \\
{\tt tauola/TauSpinner/src/tau\_reweight\_lib.cxx}

\begin{flushleft}
{\bf Calculation of event weights }
\end{flushleft}

Spin correlations matrix  $R_{ij}$  (coded in {\tt fortran})  is used by weight calculating
method (coded in {\tt C++}).

The elements of the {\tt Rij} matrix are calculated by routines {\tt dipolqq\_ } and {\tt
 dipolgamma\_ } respectively for antiquark-quark  and $\gamma\gamma$ processes.
They are provided in file {\tt TAUOLA/TauSpinner/src/initwksw.f }.
In addition to calculation,  the frame re-orientation is provided in  these 
{\tt FORTRAN } interfacing routines (Eq.~(\ref{eq:frames})). Also   
the change of index convention from {\tt FORTRAN} 1,2,3,4 to {\tt C++ } 0,1,2,3 is introduced.
Finally, minus sign originating from
$\tau^+$ V+A coupling instead of $\tau^-$ V-A is  introduced there as well.
That explain also  sign change in $R_{yy}$ versus publication~\cite{Banerjee:2023qjc}.

The function
{\tt  dipolqq\_(iqed,E,theta,channel,Amz0,Gamz0,sin2W0,alphaQED,ReA0,ImA0, ReB0,\\ImB0, ReX0,ImX0,ReY0,ImY0, GSWr0,GSWi0,Rij)}
has several input parameters and output normalised matrix {\tt Rij}, i.e. $r_{ij}=R_{ij}/R_{tt}$.
The \ {\tt E, theta} denote respectively the energy of the scattering antiquark in the $ 2 \to 2$ parton level centre-of-mass
frame and scattering angle of the outgoing $\tau$ lepton with respect to  antiquark beam direction, also in this frame.
The {\tt channel} denote type of flavour of incoming  partons.
The {\tt  A0, B0} denote anomalous magnetic and electric moments. The {\tt  X0, Y0} denote weak anomalous magnetic and electric  moments.
The {\tt iqed} flag, switches OFF/ON contribution to {\tt A0} from   the Standard Model  
$A(0)_{SM}= 1.17721(5) \times 10^{-3}$
of  the anomalous magnetic dipole moment, as calculated in ~\cite{Eidelman:2007sb}.
The {\tt Amz0,Gamz0,sin2W0,alphaQED} denotes EW parameters at Born level in $\alpha(0)$ scheme and
{\tt  GSWr0, GSWi0} denote EW form factors used for Improved Born Approximation. 

The function {\tt dipolgamma\_(iqed, E, theta, A0, B0, Rij)} has several input
parameters  {\tt iqed, E, theta, A0, B0} and output normalised matrix {\tt Rij}, i.e. $r_{ij}=R_{ij}/R_{tt}$.
The \ {\tt E, theta} denote respectively the energy of the scattering photon in the $ 2\to 2$ parton level centre-of-mass
frame and scattering angle of the outgoing $\tau$ lepton with respect to photon beam direction, also in this frame.
The {\tt  A0, B0} denote anomalous magnetic and electric moments. The {\tt iqed} flag, switches OFF/ON 
contribution to {\tt A0} from the Standard Model  $A(0)_{SM}= 1.17721(5) \times 10^{-3}$  of  the anomalous magnetic dipole moment~\cite{Eidelman:2007sb}.

For each event, which is  read from the {\tt input\_file} by:
\begin{verbatim}
int status = readParticlesFromTAUOLA_HepMC(input_file, X, tau, tau2, 
             tau_daughters, tau_daughters2);
\end{verbatim}
calculation of spin weight $WTspin$ and the corresponding relative change to the cross-section $WTprod$,
due to introduced NP models can be invoked.
\begin{verbatim}
double WTspin = calculateWeightFromParticlesH(X, tau, tau2, 
                                              tau_daughters,tau_daughters2);
double WTprod = getWtNonSM();
\end{verbatim}

By definition, always an average of spin weight $ \langle \, WTspin \, \rangle= 1.0$. To quantify the impact on particular kinematical distribution of including spin correlations for a given (SM+NP) model with respect to one used
in generated sample,  one should use product $ WTspin \cdot WTprod$ when filing histograms.
It means that only relative change in the cross-section, not the absolute one, can be accessed with
present implementation in {\tt TauSpinner}.

\begin{flushleft}
{\bf Accessing internal variables }
\end{flushleft}

Several functions ({\it getters})  are available to access internal variables  are prepared.
In particular for  each event,  with the methods
\begin{verbatim}
getZgamParametersTR(Rxx, Ryy, Rxy, Ryx);
getZgamParametersL(Rzx, Rzy, Rzz, Rtx, Rty, Rtz);
\end{verbatim}
the set of normalised $r_{ij}$ matrix components can be accessed, which might be of interest
for monitoring purposes.


\vspace{1cm}
\bibliographystyle{utphys_spires}
\bibliography{SpinZtautau_v2}

\end{document}